\documentclass[showkeys,showpacs,preprintnumbers,amsmath,amssymb]{revtex4}

\usepackage[dvipdfm]{graphicx}
\begin{document}
\title{Statistical mechanics of digital halftoning}
\author{Jun-ichi Inoue$^{1}$}
\email[e-mail: ]{j_inoue@complex.eng.hokudai.ac.jp}
\author{Yohei Saika$^2$}
\email[e-mail: ]{saika@wakayama-nct.ac.jp}
\author{Masato Okada$^3$}
\email[e-mail: ]{okada@k.u-tokyo.ac.jp}
\affiliation{
$^1$Complex Systems Engineering, Graduate School 
of Information Science and Technology, Hokkaido University, 
N14-W9, Kita-ku, Sapporo 060-0814, Japan \\
$^2$ Department of Electrical and Computer Engineering,  
Wakayama National College of Technology, 
Nada-cho, Noshima 77, Gobo-shi, Wakayama 644-0023, Japan \\
$^3$ 
Division of Transdisciplinary Science, 
Graduate School of Frontier Science, 
The University of Tokyo, 5-1-5 Kashiwanoha, Kashiwa-shi, 
Chiba 277-8561, Japan
}
\begin{abstract}
We consider the problem of 
digital halftoning from the view point of statistical mechanics. The digital halftoning 
is a sort of image processing, namely, representing each grayscale in terms of 
black and white binary dots.
The digital halftoning is achieved by making use of the threshold mask, 
namely, for each pixel, the halftoned binary 
pixel is determined as black if the original grayscale pixel 
is greater than or equal to the mask value and is determined 
as white vice versa. 
To determine the optimal value of the mask on each pixel 
for a given original grayscale image, we first assume 
that the human-eyes might recognize the black and white 
binary halftoned image as the corresponding grayscale 
one by linear filters. The Hamiltonian is 
constructed as a distance between the original and 
the recognized images which is written in terms of 
the threshold mask. 
We are confirmed that the 
system described by the Hamiltonian is regarded as 
a kind of antiferromagnetic Ising model with quenched disorders.  
By searching the 
ground state of the Hamiltonian, we obtain the optimal 
threshold mask and the resultant halftoned binary 
dots simultaneously. From the power-spectrum analysis, 
we find that the resultant binary dots image is 
physiologically plausible from the view point of human-eyes modulation properties. 
We also propose a theoretical framework to investigate statistical performance of 
inverse digital halftoning, that is, the inverse process of halftoning.
The inverse-halftoning is regarded as a special example of 
image restoration in which one should infer the original 
grayscale image from the less informative black and white binary dots. 
From the Bayesian inference view point, we rigorously show that the 
Bayes-optimal inverse-halftoning 
is achieved on a specific condition which is very similar to the so-called 
Nishimori line in the research field of spin glasses. 
Finally we show that 
both halftoning and the inverse-halftoning processes are unified under a single Hamiltonian, namely, 
it is possible for us to obtain the threshold mask, 
the halftoned and inverse-halftoned  images simultaneously by finding 
the ground state of the spin systems. 
\end{abstract}
\maketitle
\section{Introduction}
\label{sec:intro}
Recently, a lot of problems of 
information science and technology have been investigated 
by several useful tools developed in the research field 
of statistical mechanics of spin glasses \cite{Mezard,Nishi,Bishop,Mezard2009}. 
Statistical mechanics of 
information is now widely spreading in various 
subjects such as neural networks \cite{Bishop}, 
error-correcting codes \cite{Sourlas,Rujan,KabashimaSaad}, 
CDMA multi-user demodulator \cite{TanakaT}, 
image processing, {\it etc} \cite{Bishop,TanakaK}. 
Especially, image restoration by making use of 
a graphical model of Markov random fields has been 
investigated extensively from both analytical 
and numerical point of views\cite{TanakaK}. 
In the research field of image processing, 
digital halftoning instead of image restoration, 
which is defined as a process of 
generating a pattern of pixels with limited number of colors, 
especially converting a grayscale image into the binary black 
and white picture,  has been widely used in various 
practical situations in media, such as the printing 
of newspapers, fax machines and so forth \cite{Lau,Ulchney}. 

To achieve the digital halftoning, one needs the strategy to arrange 
the geometrical-combination of black and white pixels 
so as to make human-eyes to have a kind of 
{\it optical illusion}. Namely, the halftoning relies on the fact 
that the human-eyes act as a spatial low-pass filter and 
can not recognize 
any specified structure in 
the part of image dominated by high frequency components.
 
\begin{figure}[ht]
\begin{center}
\includegraphics[width=10cm]{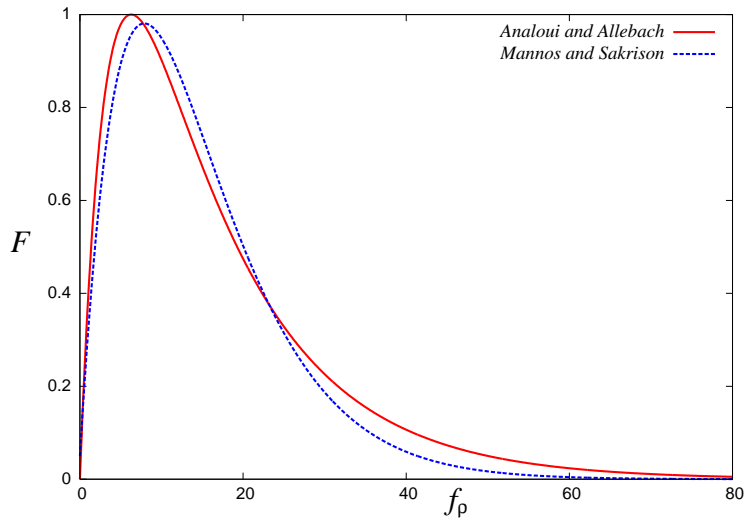}
\end{center}
\caption{\footnotesize 
Human-eye modulation transfer functions estimated by 
Analoui and Allebach \cite{Analoui} (the solid line),  and 
Mannos and Sakrison \cite{Mannos}(the broken line).}
\label{fig:fg1}
\end{figure}

Actually, to justify this fact, several 
authors proposed or estimated the so-called 
the contrast sensitivity function, or the modulation 
transfer function of human visual systems independently. 
For instance, Analoui and Allebach \cite{Analoui} 
introduced the contrast sensitivity function having the following form: 
\begin{eqnarray}
F(f_{\rho}) & =&  
k 
\left\{
\exp(-2\pi \alpha f_{\rho})
-\exp(-2\pi \beta f_{\rho})
\right\}
\label{eq:Analoui}
\end{eqnarray}
where $f_{\rho}$ denotes frequency and 
$\alpha$ and $\beta$ are constants setting to 
$0.012$ and $0.046$ respectively. 
They set the scale parameter 
$k$ to satisfy the condition 
$\max_{f_{\rho}}
F(f_{\rho})=1$, 
namely, 
$k=[(\alpha/\beta)^{-\alpha/(\alpha-\beta)}
-
(\alpha/\beta)^{-\beta/(\alpha-\beta)}]^{-1} 
\simeq 2.173938$. 
As another model of 
the human visual system, Mannos and Sakrison \cite{Mannos} 
estimated the function as 
\begin{eqnarray}
F(f_{\rho}) & = &  
2.6 
(0.0192 + 0.114f_{\rho}) 
\exp
\left[
-(0.114f_{\rho})^{1.1}
\right]
\label{eq:Mannos}.
\end{eqnarray}
We plot the shape of these contrast 
sensitivity functions (\ref{eq:Analoui}) and 
(\ref{eq:Mannos}) in FIG.\ref{fig:fg1}.
From this figure, we find that in high-frequency 
regime $f_{\rho} \gg 1$, the both contrast sensitivity functions 
decrease to zero, which means that human eyes 
cannot recognize high-frequency components in images. 
In other words, the halftone algorithm should be 
constructed such that the halftoned binary dots 
contain relatively high-frequency components 
to describe the original grayscale revels. 
This fact is an important guide for us to consider 
halftoning algorithms from the 
view point of physiology. 

Up to now, to achieve fine qualities 
of the digital halftoning, a lot of techniques, 
for instance, {\it clustered-dot} 
{\it ordered dither method}, 
{\it threshold mask method} \cite{Lau,Ulchney} 
and {\it blue noise mask method} \cite{Mista,Ulichney2}, 
{\it error-diffusion method} \cite{Floyd}, {\it etc}. have been 
proposed and developed by many researchers.
However, no attempt has been done 
to formulate the important problem 
from the view point of statistical mechanics 
of information. 
It is not difficult for us to assume that 
each pixel for the halftone and the grayscale images 
is represented by Ising spin and 
Q-Ising spin (or Potts spin), respectively. 
Therefore, it seems that 
statistical mechanical 
approach is very useful for the problem. 

On the other hand, to retrieve the original grayscale 
image from the halftone binary dots, the so-called 
{\it inverse-halftoning} has been also developed especially 
in scanner technology \cite{Inoue2007,Saika2009}. As such an inverse halftoning strategy, 
the conventional smoothing filters have been used widely.  
However, it is worth while for us to 
look for the alternative or reconsider the problem 
from the view point of statistical mechanics of 
disordered spin systems. 

In this paper, we propose a statistical-mechanical modeling 
of both digital halftoning and the inverse process, 
namely, inverse-halftoning. To attempt to generate the halftoned binary dots, we first introduce a threshold 
mask matrix with the same size as the original image. For the threshold mask, 
each pixel of the halftoned image is determined 
as a black pixel if the value of the original grayscale 
pixel is greater than the component of the threshold mask 
matrix and is determined as a white pixel vice versa. 
Then, we assume that human eyes might recognize 
the original grayscale image by a linear filter. 
By taking into account that the distance between 
the original and the recognized images is written 
in terms of the threshold mask, we naturally 
introduce a Hamiltonian to be minimized to 
determine the optimal threshold mask as the ground state. 
We explicitly show that 
the system described by the Hamiltonian is a variant of 
the antiferromagnetic Ising model with disorders. 
For a demonstration of our method to generate 
the optimal threshold mask, 
computer simulations of simulated annealing \cite{Kirkpatrick,Geman}
is carried out. To investigate the statistics 
of the threshold mask image and the halftoned image, 
we evaluate the power-spectrum numerically. 
 
On the other hand, in order to investigate statistical 
performance of the inverse digital halftoning,  
in our previous study \cite{Inoue2007}, we introduced the standard regularization theory for 
a Markov random-fields model 
to represent original grayscale image and construct 
the inverse process as a kind of 
image reconstruction of the grayscale image 
from a given halftone binary dots. 
Then, we evaluated the statistical performance 
by making use of Markov chain Monte Carlo (MCMC) 
simulations and analysis of the infinite-range mean-field model. 
We also investigated the inverse-halftoning 
process as a dynamics of disordered spin systems. 
In this paper, we reconsider the 
inverse halftoning problem 
from the Bayesian inference point of view. 
We show that 
the Bayes-optimal inverse-halftoning 
that minimizes the mean square error 
is achieved on a specific condition 
which is similar to the so-called Nishimori line \cite{Nishi1981} 
in the research field of spin glasses. 
Finally we discuss the possibility 
that both halftoning and the inverse-halftoning processes are unified under a single Hamiltonian, namely, 
we argue whether it is possible for us to obtain the threshold mask, 
the halftoned and inverse-halftoned  images simultaneously by finding 
the ground state of the spin systems or not.

This paper is organized as follows. 
In the next Sec. \ref{sec:model}, we explain our procedure 
to generate the optimal threshold mask for 
a given original grayscale image. In this section, 
the Hamiltonian is introduced as a function of 
the threshold mask under the assumption that 
the human-eyes might recognize the halftoned image 
as the corresponding original grayscale version 
by a linear filter. In Sec. \ref{sec:Numerical}, 
we solve the combinatorial optimization problem 
described by the Hamiltonian by making 
use of the simulated annealing. 
We show the resulting halftoned image 
for a standard grayscale image. 
We find that our algorithm generates the binary 
dots which induce the illusion of a continuous-tone image. 
We also investigate the power-spectrum statistics 
of the threshold mask array and the halftoned image. 
In Sec.\ref{sec:INV}, we formulate our 
inverse-halftoning procedure based on the MPM estimation of 
Bayesian statistics. We show clearly that the Bayes-optimal inverse-halftoning 
that minimizes the mean-square error 
is achieved on a specific condition. 
In Sec. \ref{sec:simult}, we discuss the possibility 
that both halftoning and the inverse-halftoning processes are unified under a single Hamiltonian.  
The last section is summary. 
\section{Statistical-mechanical modeling of halftoning}
\label{sec:model}
We first deal with the `forward problem' of 
the digital halftoning. 
In this section, we first explain the procedure to generate 
the binary dots to represent the original grayscale image 
via what we call {\it threshold dither method} \cite{Lau,Ulchney}. 
Then, we explain how one formulates the halftoning 
processes as a problem of disordered spin systems and 
why statistical mechanics of 
information is useful for the problem. 
\subsection{Halftoning by threshold dither method}
Let us first define the original 
grayscale image located on the square lattice with size 
$L_{1} \times L_{2}$ by 
\begin{eqnarray}
\mbox{\boldmath $g$} & = & 
\left\{
g_{x,y} \in 
0,1,\cdots,Q-1|
x=1,2,\cdots,L_{1},y=1,2,\cdots,L_{2}
\right\}.
\label{eq:OriginalImage}
\end{eqnarray}
To generate the binary dots to represent 
the grayscale image 
$\mbox{\boldmath $g$}$, we introduce the threshold 
mask matrix with the size $l_{1} \times l_{2}$ 
($l_{1} \leq L_{1}, l_{2} \leq L_{2}$): 
\begin{eqnarray}
\mbox{\boldmath $t$} & = & 
\left\{
t_{x,y} \in 
0,1,\cdots,Q-1|
x=1,2,\cdots,l_{1}, 
y=1,2,\cdots,l_{2}
\right\}
\label{eq:MaskImage}
\end{eqnarray}
We should keep in mind that 
each component of the 
matrix $t_{x,y}$ should be satisfied the 
following periodic boundary condition: 
\begin{equation}
t_{x,y} = 
t_{x \pm l_{1},y}=t_{x,y \pm l_{2}},\,\,\,\,
L_{1} \equiv 0 \,\,\,({\rm mod} \,\,l_{1}),\,\,
L_{2} \equiv 0 \,\,\,({\rm mod} \,\,l_{2}). 
\label{eq:boundary_matM}
\end{equation}
Then, the halftoned image is defined by 
\begin{eqnarray}
\mbox{\boldmath $h$} & = & 
\left\{
h_{x,y} \in 0,1|
x=1,2,\cdots,L_{1}, y=1,2,\cdots,L_{2}
\right\}
\label{eq:HalftoneImage}
\end{eqnarray}
and each pixel in the 
$\mbox{\boldmath $h$}$ is calculated as 
\begin{eqnarray}
\forall_{x,y}\,\,\,\,\,\,
h_{x,y} & = & 
\Theta (g_{x,y}-t_{x,y})
\label{eq:trans_to_half},
\end{eqnarray}
where the step function 
$\Theta (x)$ is defined conventionally as 
\begin{eqnarray}
\Theta (x) & = & 
\left\{
\begin{array}{cc}
1 & (x \geq 0) \\
0 & (x < 0) 
\end{array}
\right..
\end{eqnarray}
For this setup of the halftoning procedure, 
a pixel of the resulting halftoned binary image is set 
to one if the pixel of the original 
grayscale image is greater than or equal 
to the corresponding pixel of the threshold mask; 
otherwise the pixel is set to zero. 

\begin{figure}[ht]
\begin{center}
\includegraphics[width=5.8cm]{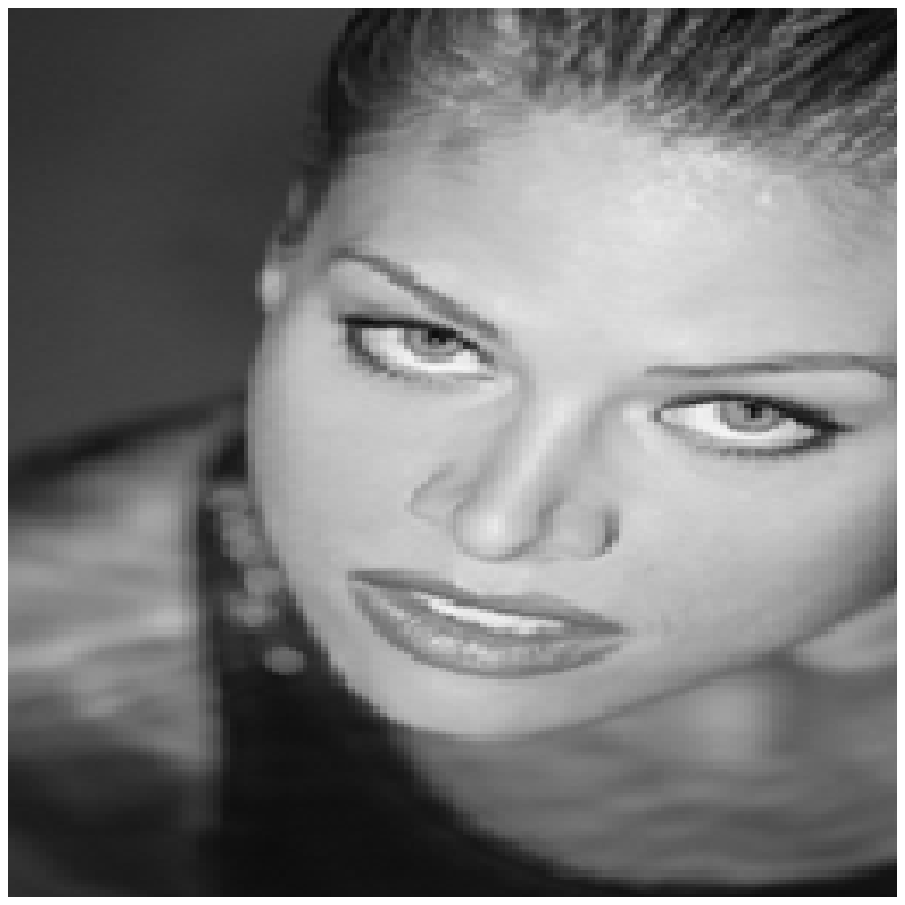} 
\includegraphics[width=5.8cm]{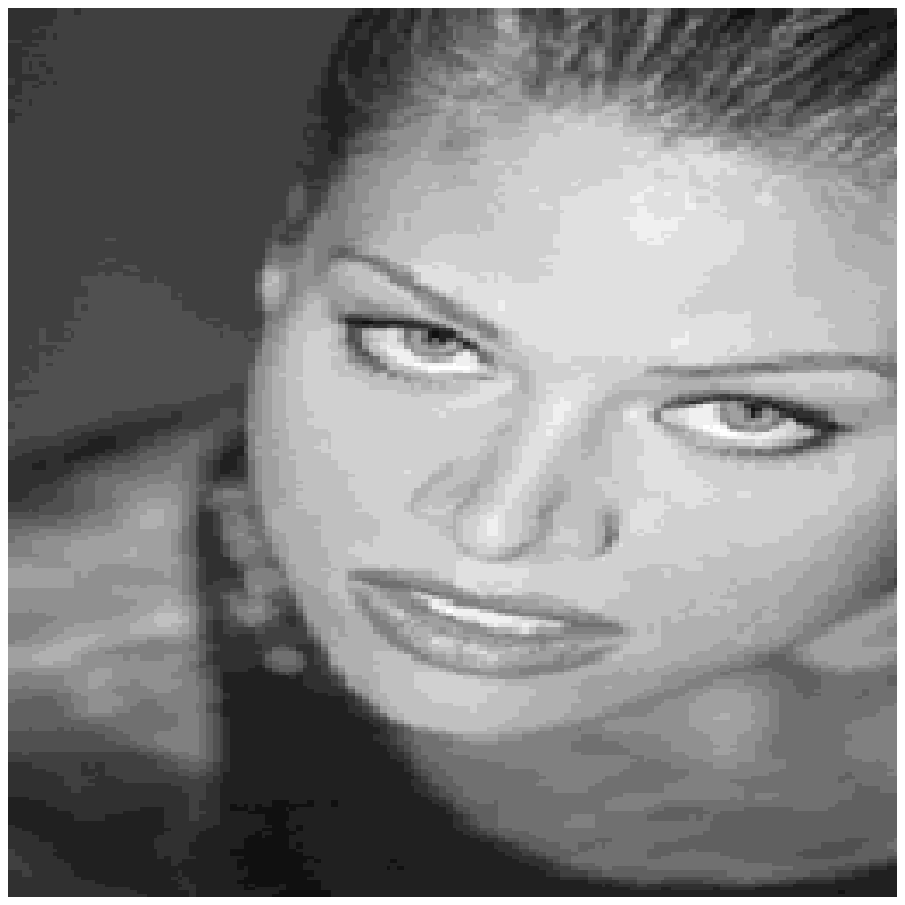} 
\includegraphics[width=5.8cm]{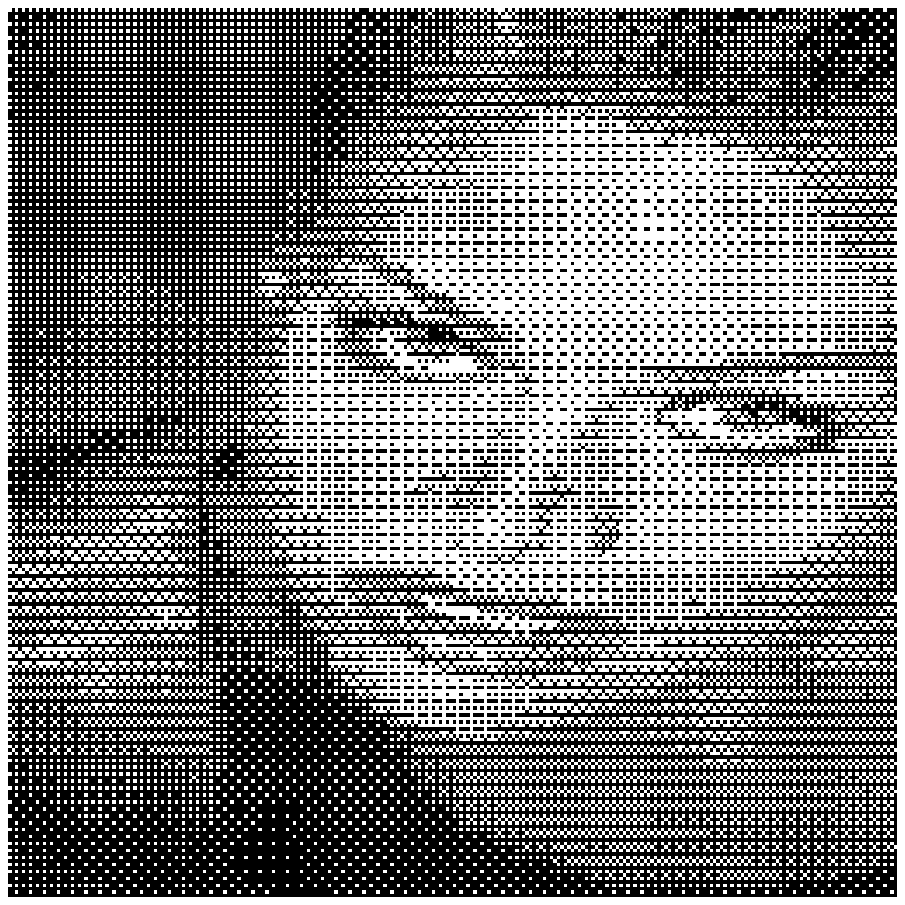} 
\end{center}
\caption{\footnotesize 
An example of 
digital halftoning by using 
the Bayers' matrix (\ref{eq:Bayer1}). 
The $256$-grayscale image (left) and 
the original images in which 
the grayscale levels are reduced from 
$256$ to $16$ (center). 
The right panel shows the halftoned image 
obtained by the dither method of 
(\ref{eq:Bayer1}). 
We set $Q=16$ and $L_{1} \times L_{2}=256 \times 256$.}
\label{fig:fg01}
\end{figure}

\begin{figure}[ht]
\begin{center}
\includegraphics[width=5.8cm]{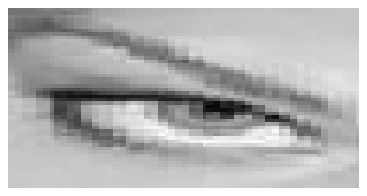} 
\includegraphics[width=5.8cm]{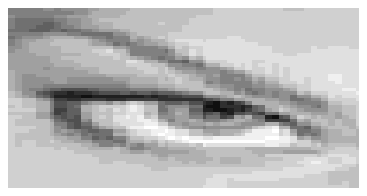} 
\includegraphics[width=5.8cm]{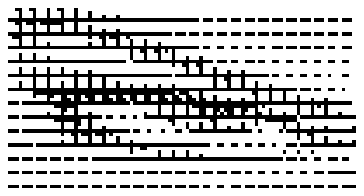} 
\end{center}
\caption{\footnotesize 
The result focusing on a tiny part (the `right eye' of the woman) of the same image as in FIG. \ref{fig:fg01}. 
}
\label{fig:fg00}
\end{figure}

For instance, the digital halftoning is achieved 
by threshold mask of Bayers' type for 
$Q=16$ with size $l_{1} \times l_{2}=4 \times 4$:  
\begin{eqnarray}
\left(
\begin{array}{cccc}
t_{1,1} & t_{1,2} & t_{1,3} & t_{1,4} \\
t_{2,1} & t_{2,2} & t_{2,3} & t_{2,4} \\
t_{3,1} & t_{3,2} & t_{3,3} & t_{3,4} \\
t_{4,1} & t_{4,2} & t_{4,3} & t_{4,4} 
\end{array}
\right) & = & 
\left(
\begin{array}{cccc}
0 & 8 & 2 & 10 \\
12 & 4 & 14 & 6 \\
3 & 11 & 1 & 9 \\
15 & 7 & 13 & 5 
\end{array}
\right)
\label{eq:Bayer1}
\end{eqnarray}
Namely, for a part with 
size $4 \times 4$ of the 
original image $\mbox{\boldmath $g$}$, say, for 
\begin{eqnarray}
\left(
\begin{array}{cccc}
g_{1,1} & g_{1,2} & g_{1,3} & g_{1,4} \\
g_{2,1} & g_{2,2} & g_{2,3} & g_{2,4} \\
g_{3,1} & g_{3,2} & g_{3,3} & g_{3,4} \\
g_{4,1} & g_{4,2} & g_{4,3} & g_{4,4} 
\end{array}
\right) & = & 
\left(
\begin{array}{cccc}
5 & 5 & 5 & 5 \\
5 & 5 & 4 & 4 \\
4 & 4 & 4 & 4 \\
4 & 4 & 4 & 4 
\end{array}
\right), 
\end{eqnarray}
we have the corresponding 
block of the halftone image $\mbox{\boldmath $h$}$ as 
\begin{eqnarray}
\left(
\begin{array}{cccc}
h_{1,1} & h_{1,2} & h_{1,3} & h_{1,4} \\
h_{2,1} & h_{2,2} & h_{2,3} & h_{2,4} \\
h_{3,1} & h_{3,2} & h_{3,3} & h_{3,4} \\
h_{4,1} & h_{4,2} & h_{4,3} & h_{4,4} 
\end{array}
\right) & = &  
\left(
\begin{array}{cccc}
\Theta(g_{1,1}-t_{1,1}) & 
\Theta(g_{1,2}-t_{1,2}) & 
\Theta(g_{1,3}-t_{1,3}) & 
\Theta(g_{1,4}-t_{1,4}) \\
\Theta(g_{2,1}-t_{2,1}) & 
\Theta(g_{2,2}-t_{2,2}) & 
\Theta(g_{2,3}-t_{2,3}) & 
\Theta(g_{2,4}-t_{2,4}) \\
\Theta(g_{3,1}-t_{3,1}) & 
\Theta(g_{3,2}-t_{3,2}) & 
\Theta(g_{3,3}-t_{3,3}) & 
\Theta(g_{3,4}-t_{3,4}) \\
\Theta(g_{4,1}-t_{4,1}) & 
\Theta(g_{4,2}-t_{4,2}) & 
\Theta(g_{4,3}-t_{4,3}) & 
\Theta(g_{4,4}-t_{4,4}) 
\end{array}
\right) = 
\left( 
\begin{array}{cccc}
1 & 0 & 1 & 0 \\
0 & 1 & 0 & 0 \\
1 & 0 & 1 & 0 \\
0 & 0 & 0 & 0 
\end{array}
\right).
\end{eqnarray}
In FIG. \ref{fig:fg01}, 
we show a typical 
example of the original 
grayscale image ($256$ grayscale levels: left panel),  
the image with reduced grayscales $Q=16$: center panel) and 
the resulting halftone image 
obtained by the Bayer's 
threshold mask (\ref{eq:Bayer1}).  
The size of the image is 
$L_{1} \times L_{2}=256 \times 256$ and the number of the 
grayscale 
levels is $256$. 
Here we first converted the original 
$256$-grayscales 
$\mbox{\boldmath $g$}^{'}$ 
image  to  the reduced $|\mathbb{N}(x,y)|=Q (< 256)$-grayscale 
image $\mbox{\boldmath $g$}$ 
by the following transform: 
\begin{eqnarray}
C(x:Q) & = & 
\sum_{k=0}^{Q-1}
k 
\left\{
\Theta 
\left(
x-
\frac{256}{Q}
(k+1)
\right)
-\Theta 
\left(
x- 
\frac{256}{Q}k
\right)
\right\}
\label{eq:transf}
\end{eqnarray}
and used 
$\forall_{x,y} \,\,\,
g_{x,y}=C(g^{'}_{x,y}:Q)$ as the original 
pixel in (\ref{eq:trans_to_half}). 
From the right panel of this figure, we find that 
the grayscale-levels of the image in the middle panel 
is described by black and white binary dots. 
To see the detail of the binary dots which 
represent the grayscale levels in the original image, 
in FIG. \ref{fig:fg00},  
we show the result focusing on a tiny part (the `right eye' of the woman) of the same image in FIG. \ref{fig:fg01}. 
\mbox{} 

Therefore, once we obtain the appropriately  
threshold matrix 
$\mbox{\boldmath $t$}$, 
the halftone binary dots 
$\mbox{\boldmath $h$}$ is 
determined uniquely, 
and for this reason, 
the quality of 
the halftoning is 
dependent on how we choose the 
threshold mask appropriately. 
In this paper, 
we consider the case of 
the threshold mask 
having the same size as that of 
the original grayscale image, 
that is, 
the case of $l_{1}=L_{1}, l_{2}=L_{2}$, and 
propose an algorithm to obtain 
the optimal threshold mask 
as lowest energy states of 
the Hamiltonian which 
is a square distance 
between 
the original and 
the recognized images. We shall 
introduce it in the next subsection. 
\subsection{A recognition model of human vision}
As we explained in the previous subsection, 
our main problem is now to determine the threshold mask 
$\mbox{\boldmath $t$}$ efficiently. 
For this purpose, we might assume that human 
eyes might retrieve the grayscale image, 
let us call, the recognized image: 
\begin{eqnarray}
\mbox{\boldmath $s$} & = & 
\left\{
s_{x,y} \in 0,1,\cdots, Q-1|
x=1,2,\cdots,L_{1},y=1,2,\cdots,L_{2}
\right\}
\end{eqnarray}
from the halftoned image 
$\mbox{\boldmath $h$}$ by making 
use of the following linear filter: 
\begin{eqnarray}
\forall_{x,y} \,\,\,\,\,
s_{x,y} & = &  
\sum_{i,j \in \mathbb{N} (x,y)}
W_{x-i,y-j}\,h_{i,j}
\label{eq:def_filter}, 
\end{eqnarray}
where $\mathbb{N}(x,y)$ 
denotes the nearest neighboring 
pixels of the pixel 
located at $(x,y)$ and the point $(x,y)$ itself. 
Apparently, for a two-dimensional 
square lattice, the ingredients of 
$\mathbb{N}(x,y)$ are 
$(x,y),(x-1,y+1),(x,y+1),
(x+1,y+1),(x-1,y),(x+1,y),
(x-1,y-1),(x,y-1),(x+1,y-1)$ 
and the size is 
$|\mathbb{N} (x,y)|=9$. 
The above choice for 
the recognition model 
comes from our assumption 
that human-eyes might recognize a part of 
halftone pictures more dark when the density 
of black pixels 
located on the part is relatively high. The justification of 
this simple assumption 
could be checked by comparing 
the recognized image 
$\mbox{\boldmath $s$}$, 
which is calculated in terms of 
the resultant halftone binary dots 
via (\ref{eq:def_filter}), with 
the original grayscale image 
$\mbox{\boldmath $g$}$. 
We shall discuss the results in Sec. \ref{sec:Numerical}. 

Obviously, for the simplest choice, 
we might set the weight $W_{x,y}$  
appearing in (\ref{eq:def_filter}) 
as $\forall_{x,y}\,\,\,W_{x,y}=1$. 
We should notice that 
for this choice, the spatial structure 
of the black dots in 
$\mathbb{N}(x,y)$ is not taken into account and 
the number $|\mathbb{N}(x,y)|$ itself determines 
the grayscale level of the corresponding 
pixel $s_{x,y}$ in the  
recognized image $\mbox{\boldmath $s$}$. 
Therefore, the choice of the effective grayscale 
level can be controlled by choosing the number of 
the nearest neighboring pixels $|\mathbb{N} (x,y)|$. 
Taking into account this limitation, 
we first reduce $256$-grayscale levels to 
$Q(<256)$ by (\ref{eq:transf}) and regard the $Q$-grayscale image as the 
$\mbox{\boldmath $g$}$. 

We also might utilize the other choice such as the following 
two-dimensional Gaussian-type: 
\begin{eqnarray}
W_{x-i,y-j} & = & 
\frac{1}{2\pi 
\sqrt{{\rm det}
\mbox{\boldmath $\Sigma$}}}
\,\exp
\left[
-\frac{
\mbox{\boldmath $z$}^{T}
\mbox{\boldmath $\Sigma$}^{-1}
\mbox{\boldmath $z$}
}{2}
\right],\,\,
\mbox{\boldmath $\Sigma$}=
\left(
\begin{array}{cc}
\sigma_{x}^{2} & 
\rho\,
\sigma_{x}\sigma_{y} \\
\rho\,
\sigma_{x}\sigma_{y} & \sigma_{y}^{2}
\end{array}
\right),\,\,
\mbox{\boldmath $z$}^{T}=
(x-i,y-j). 
\end{eqnarray}
For this choice, 
not only the number $|\mathbb{N}(x,y)|$ but 
also the spatial structure 
determines the grayscale levels in 
the recognized image. 
However, 
we naturally assume that 
the spatial structure 
does not affect the quality of halftoning 
if the size of the window 
$|\mathbb{N}(x,y)|$ is 
small enough in comparison with 
the image size $L^{2}$. 

In this paper, we use the 
following specific choice of the weight $W_{x-i,x-j}$ with size $|\mathbb{N}(x,y)|=5 \times 5=25$ as 
\begin{eqnarray}
\left(
\begin{array}{lllll}
W_{x-2,y+2} & W_{x-1,y+2} & W_{x,y+2} & W_{x+1, y+2} & W_{x+2,y+2}  \\ 
W_{x-2,y+1} & W_{x-1,y+1}  & W_{x,y+1} & W_{x+1,y+1} & W_{x+2,y+1} \\
W_{x-2,y} & W_{x-1,y} & W_{x,y}  & W_{x+1,y} & W_{x+2,y} \\
W_{x-2,y-1} & W_{x-1,y-1} & W_{x,y-1} & W_{x+1,y-1} & W_{x+2,y-1} \\
W_{x-2,y-2} & W_{x-1,y-2} & W_{x,y-2} & W_{x+1,y-2} & W_{x+2,y-2}
\end{array}
\right) & = & 
\left(
\begin{array}{ccccc}
5 & 5 & 15 & 5 & 5 \\ 
5 & 10 & 15 & 10 & 5 \\
15 & 15 & 35 & 15 & 15 \\
5 & 10 & 15 & 10 & 5 \\
5 & 5 & 15 & 5 & 5  
\end{array}
\right). 
\end{eqnarray}
We should keep in mind that each 
recognized pixel $s_{x,y}$ takes the 
minimum $s_{x,y}=0$ and the maximum $s_{x,y}=255$ 
from the definition (\ref{eq:def_filter}). 
Hence, each pixel in the recognized image $\mbox{\boldmath $s$}$ 
calculated by (\ref{eq:def_filter}) 
represents `defective' $256$-grayscale levels, 
that is, $\forall_{x,y}\,\,\,s_{x,y} \in 
\{0,5,10,\cdots,255\}$.

Then, we might choose a strategy 
to determine the threshold mask 
$\mbox{\boldmath $t$}$ that minimizes 
the square distance between the original and 
the recognized images. Namely, we minimize 
the energy $\Vert 
\mbox{\boldmath $g$}-
\mbox{\boldmath $s$}
\Vert^{2} \equiv   
\sum_{x=1}^{L_{1}} 
\sum_{y=1}^{L_{2}}
(
g_{x,y}-s_{x,y})^{2}$. 
We should notice that from 
the relation (\ref{eq:def_filter}), 
the above distance $\Vert 
\mbox{\boldmath $g$}-
\mbox{\boldmath $s$}
\Vert^{2}$ 
is written in terms of 
the threshold mask $\mbox{\boldmath $t$}$ 
for a given original image 
$\mbox{\boldmath $g$}$. 
By substituting (\ref{eq:def_filter}) into 
the distance $\Vert 
\mbox{\boldmath $g$}-
\mbox{\boldmath $s$}
\Vert^{2}$, we obtain the Hamiltonian of the system as a function of the threshold mask 
$\mbox{\boldmath $t$}$. 
Thus, our digital halftoning 
is now reduced to a combinatorial 
optimization problem of the following Hamiltonian: 
\begin{eqnarray}
\mathcal{H}(\mbox{\boldmath $t$}|
\mbox{\boldmath $g$}) & = &  
\sum_{x=1}^{L_{1}}
\sum_{y=1}^{L_{2}}
\left\{
g_{x,y}-
\sum_{i,j \in 
\mathbb{N} (x,y)}
W_{x-i,y-j}
\Theta (g_{i,j}-t_{i,j})
\right\}^{2}
\label{eq:Hamiltonian}
\end{eqnarray}
The above Hamiltonian 
$\mathcal{H}(\mbox{\boldmath $t$}|
\mbox{\boldmath $g$})$ 
is a starting point 
of statistical-mechanical modeling 
of the digital halftoning. 
We should bear in mind that 
in the above Hamiltonian, 
$\mbox{\boldmath $t$}$ are 
dynamical variables, whereas, 
$\mbox{\boldmath $g$}$ are 
quenched disorders. 
As we have the Hamiltonian of the system, 
one can generate statistical ensembles 
according to standard statistical mechanics. 
In its thermal equilibrium at temperature 
$\beta^{-1}$, 
each possible microscopic state 
$\mbox{\boldmath $t$}$ 
obeys the following Boltzmann-Gibbs distribution: 
\begin{eqnarray}
P_{\beta}(\mbox{\boldmath $t$}|
\mbox{\boldmath $g$}) & = & 
\frac{
{\exp [-\beta 
\mathcal{H}(\mbox{\boldmath $t$}|
\mbox{\boldmath $g$})]}
}
{
\prod_{x=1}^{L_{1}}
\prod_{y=1}^{L_{2}}
\sum_{t_{x,y}=0}^{Q-1}
\exp [
-\beta \mathcal{H} (\mbox{\boldmath $t$}|
\mbox{\boldmath $g$})]
}
\end{eqnarray}
One of the simplest uses of 
the Hamiltonian 
is to regard its minimum energy state 
as the optimal threshold mask 
$\hat{\mbox{\boldmath $t$}} \equiv 
\{
\hat{t}_{x,y} \in 0,1|
x=1,2,\cdots,L_{1}, y=1,2,\cdots,L_{2}\}$. 
Namely, we have 
\begin{eqnarray}
\hat{
\mbox{\boldmath $t$}}& = & 
\arg\max_{\mbox{\scriptsize \boldmath $t$}} 
P_{\beta} (\mbox{\boldmath $t$}|
\mbox{\boldmath $g$}),
\end{eqnarray}
which is rewritten in terms of the 
statistical-mechanical terminology as
\begin{eqnarray}
\forall_{x,y}\,\,\,\, \hat{t}_{x,y} & = & 
\lim_{\beta \to \infty}
\prod_{x=1}^{L_{1}}
\prod_{y=1}^{L_{2}}
\sum_{t_{x,y}=0}^{Q-1}
t_{x,y}
P_{\beta} (\mbox{\boldmath $t$}|
\mbox{\boldmath $g$})
\label{eq:mapcal}.
\end{eqnarray}
Calculating the 
$Q^{L_{1} \times L_{2}}$-sums: $\prod_{x=1}^{L_{1}}
\prod_{y=1}^{L_{2}}
\sum_{t_{x,y}=0}^{Q-1}
(\cdots)$ 
and taking the zero-temperature limit are 
performed by Gibbs sampler with 
temperature annealing procedure 
during the Monte Carlo steps. 
The solution might be a good candidate 
for the suitable threshold mask. 

We should mention that our procedure is categorized 
in the method `model-based halftoning' which means 
that the threshold mask is dependent on the original image. 
In engineering, there exist several such algorithms \cite{Pappas,Pappas2,Pappas3}. However, we should stress that 
these studies are completely different from our statistical-mechanical modeling. 
\subsection{The corresponding spin system}
It is helpful for us to consider the meaning of the Hamiltonian (\ref{eq:Hamiltonian}) 
from the view point of disordered spin systems. 
To see the relationship between halftoning processes and the corresponding disordered spin system, 
we expand the square of the Hamiltonian (\ref{eq:Hamiltonian}). Then, we have 
\begin{eqnarray}
\mathcal{H}(\mbox{\boldmath $t$}|\mbox{\boldmath $g$}) & = & 
\frac{1}{2}
\sum_{x=1}^{L_{1}}
\sum_{y=1}^{L_{2}}
\sum_{(i,j) \neq (k,l) \in 
\mathbb{N}(x,y)}
\sum_{(k,l) \in 
\mathbb{N} (x,y)}
J_{ij,kl}S_{i,j}S_{k,l}
+\sum_{x=1}^{L_{1}}
\sum_{y=1}^{L_{2}}
g_{x,y}
\sum_{i,j \in 
\mathbb{N}(x,y)}
h_{i,j}S_{i,j}
\end{eqnarray}
where we defined 
$h_{i,j} \equiv W_{x-i,y-j} >0, 
J_{ij,kl} \equiv W_{x-j,y-j}W_{x-k,y-l}>0$ and 
$S_{i,j}={\rm sgn}(g_{ij}-t_{ij}) \in \{-1,+1\}$. 
It should be noted that 
we canceled a constant term 
$\sum_{x=1}^{L_{1}} 
\sum_{y=1}^{L_{2}} (g_{x,y})^{2}$ and 
used the relation $\Theta (x)=(1+{\rm sgn}(x))/2$. 

Thus, our system is nothing but 
an Ising model defined on the two-dimensional square lattice 
with antiferromagnetic interactions with random field $\mbox{\boldmath $g$}$ on pixel. 
However, the ground state 
which minimizes the above Hamiltonian 
with respect to $\mbox{\boldmath $t$}$ is 
complicated due to the quenched variables as a given original image 
$\mbox{\boldmath $g$}$ 
appearing in the argument of 
sign function such as 
${\rm sgn}(g_{ij}-t_{ij})$. 
The lowest energy state in the dynamical variable space of 
$\mbox{\boldmath $t$}$ might be highly degenerated  
and it means that we can use various techniques 
developed in the field of statistical physics 
of disordered spin systems to obtain 
the lowest energy state and to investigate 
the lowest energy properties. In the next section, 
we show some demonstrations to make the halftoned 
binary dots that imitate the original 
grayscale image by minimizing the Hamiltonian  
(\ref{eq:Hamiltonian}) via simulated annealing. 
\section{Numerical experiments}
\label{sec:Numerical}
\begin{figure}[ht]
\begin{center}
\includegraphics[width=8.5cm]{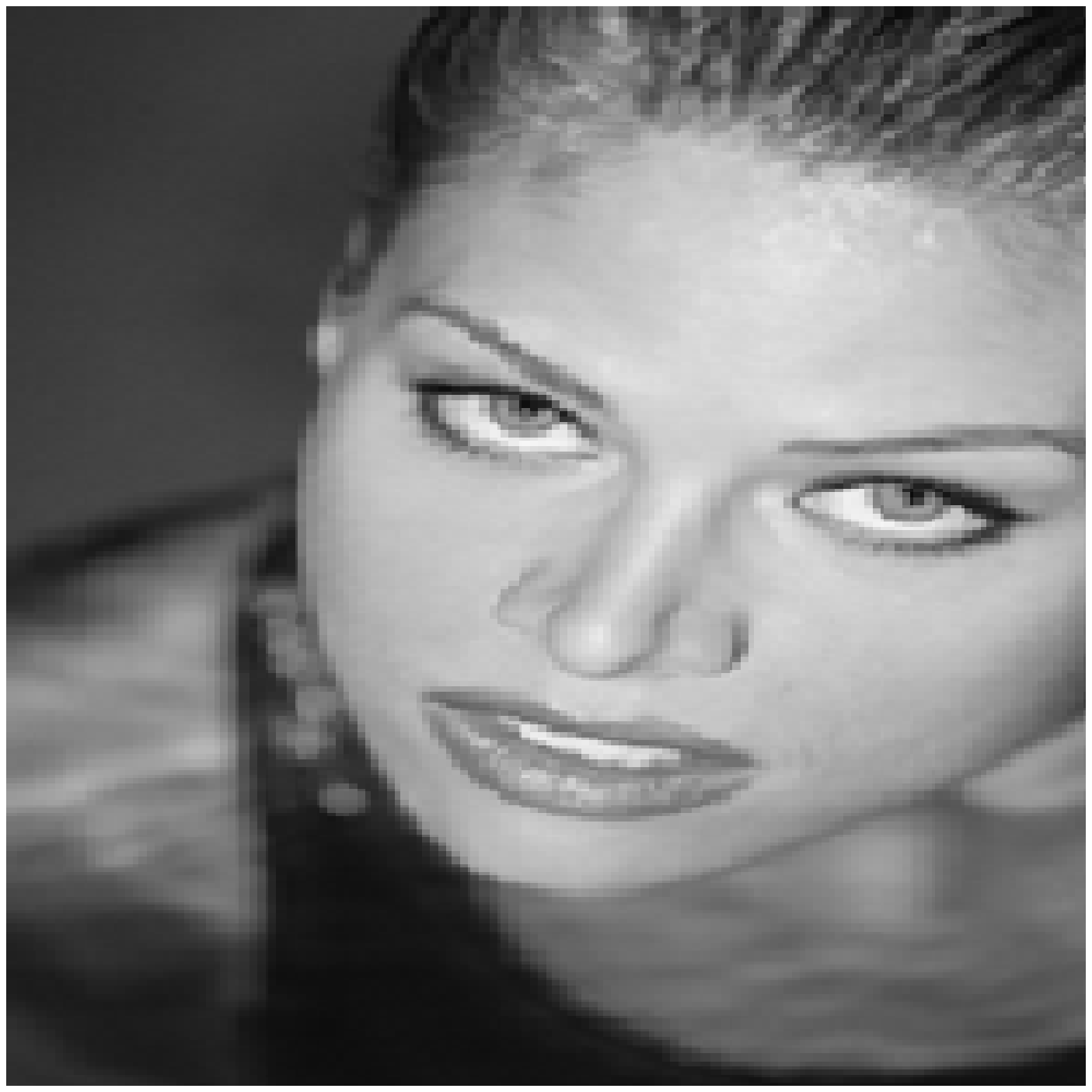}
\includegraphics[width=8.5cm]{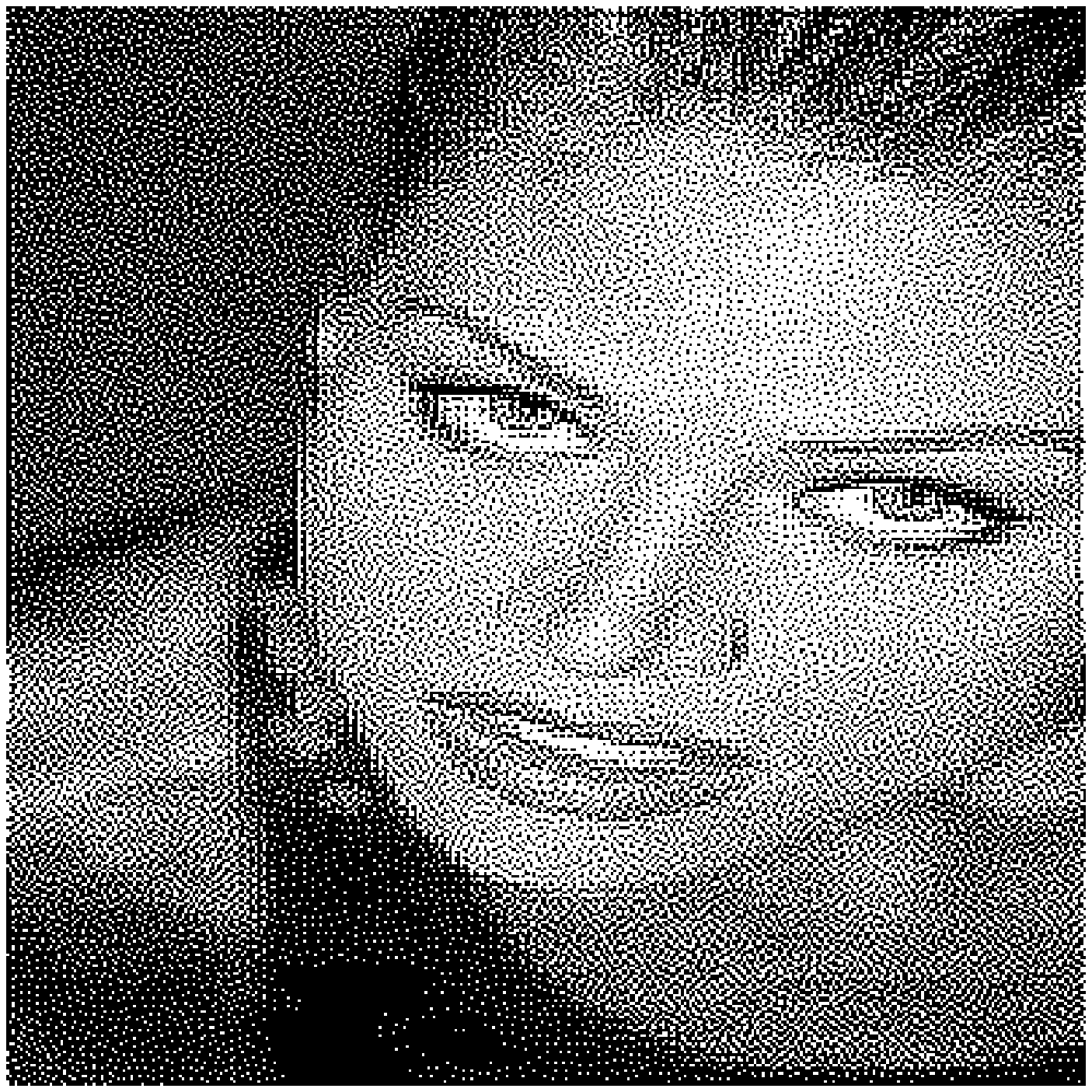}
\includegraphics[width=8.5cm]{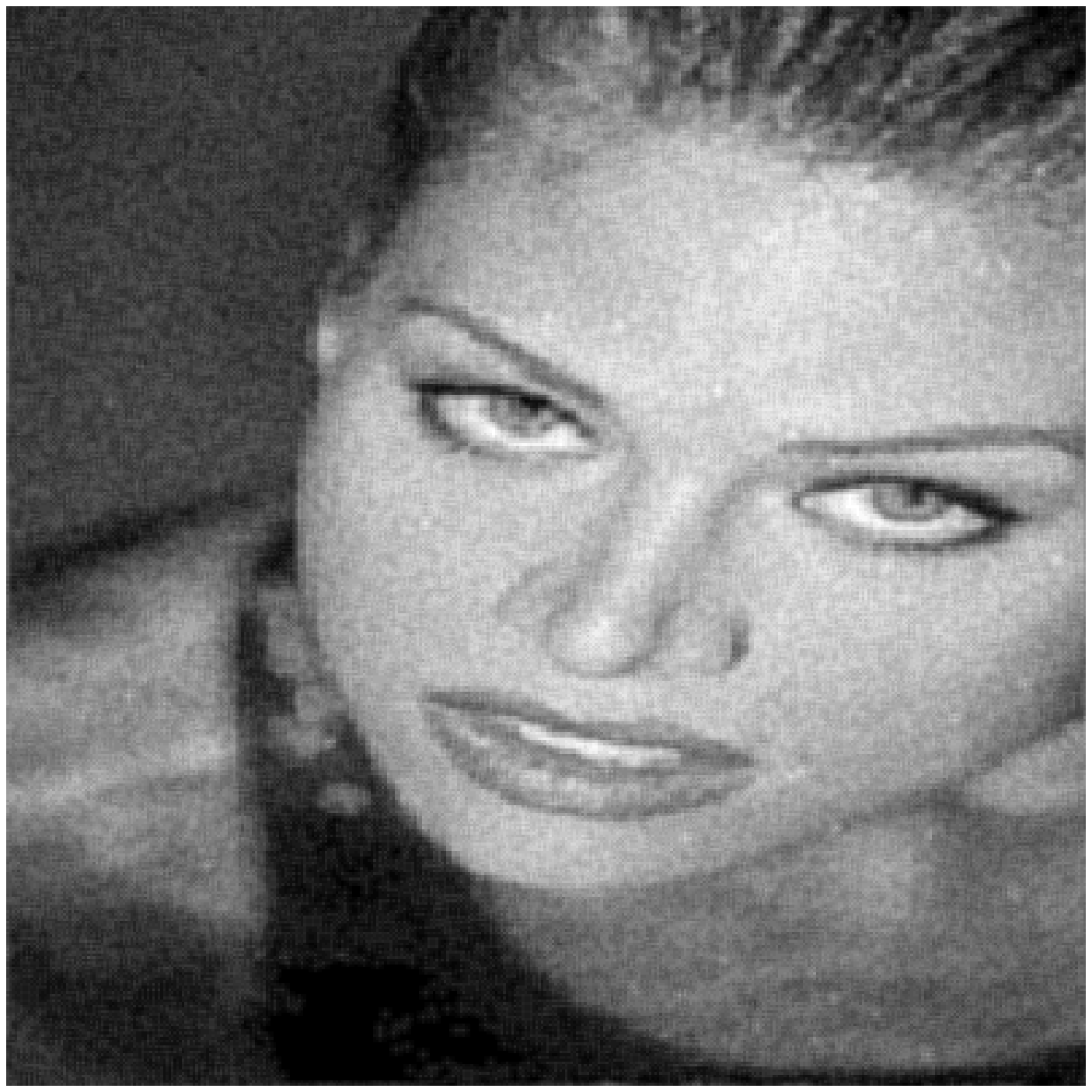}
\includegraphics[width=8.5cm]{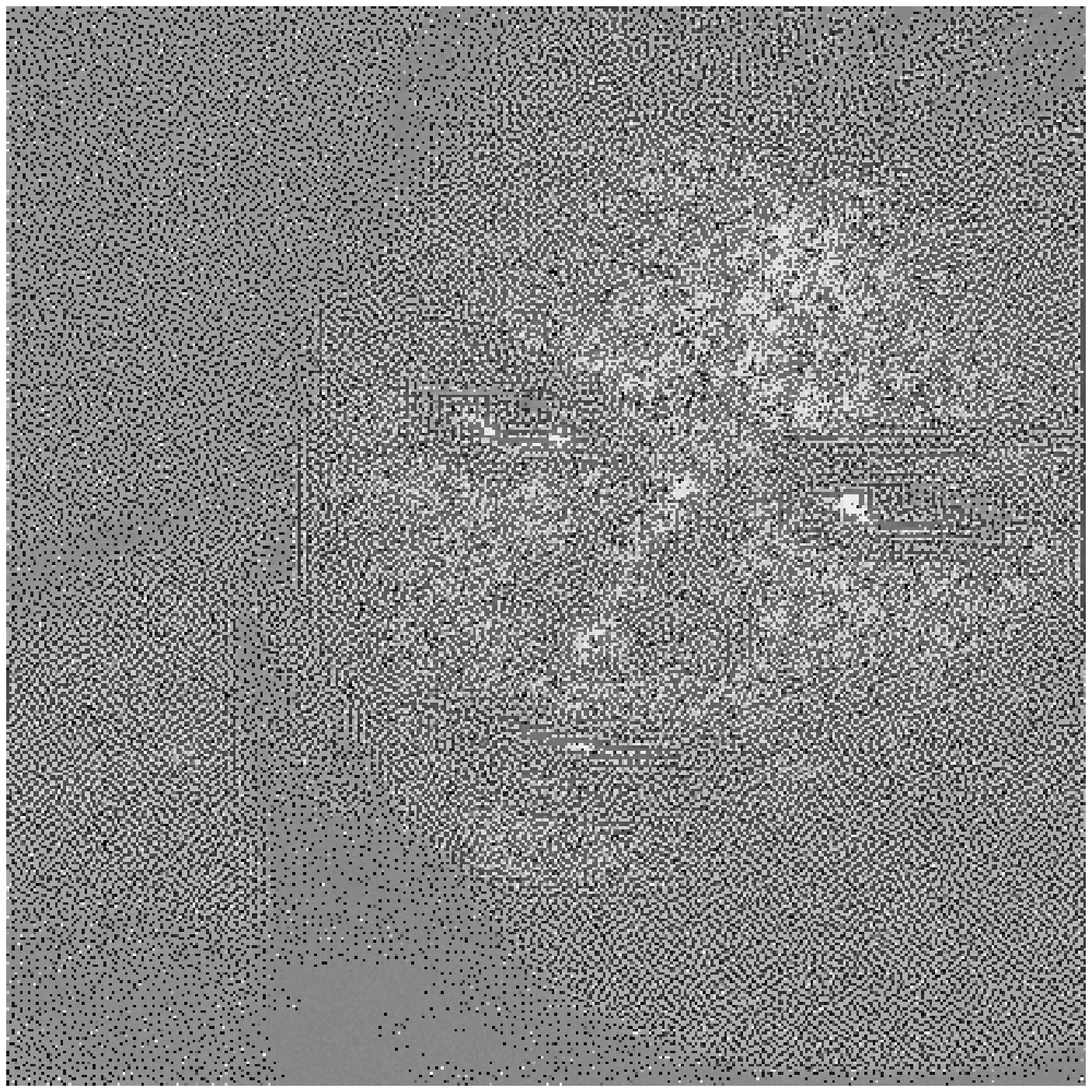}
\end{center}
\caption{\footnotesize 
From the upper left to 
the lower right, 
the original image of $256$-grayscale levels, 
the halftoned image,  
the threshold mask, 
and the recognized image obtained by our algorithm. 
The size of the images is 
$L_{1} \times L_{2} = 400 \times 400$. 
}
\label{fig:fg02}
\end{figure}

In this section, we show some 
demonstrations to generate the optimal 
threshold mask and the resulting halftoned image 
for a given standard grayscale image.  We shall demonstrate our algorithm for one of the 
well-known standard images 
shown in FIG.\ref{fig:fg02} (upper left). 
For the original image $\mbox{\boldmath $g$}$ with the size 
$L_{1} \times L_{2}=400 \times 400$ as input data, 
we construct the Hamiltonian 
$\mathcal{H}(\mbox{\boldmath $t$}|\mbox{\boldmath $g$})$ 
and carry out the calculation 
of the threshold mask $\mbox{\boldmath $\hat{t}$}$ given by (\ref{eq:mapcal}) 
via simulated annealing 
with the temperature schedule 
$\beta \sim \sqrt{t}$. 
For the solution of the threshold mask $\mbox{\boldmath $\hat{t}$}$, we make 
the halftoned image $\mbox{\boldmath $h$}$ according to (\ref{eq:HalftoneImage}).

\begin{figure}[ht]
\begin{center}
\includegraphics[width=10cm]{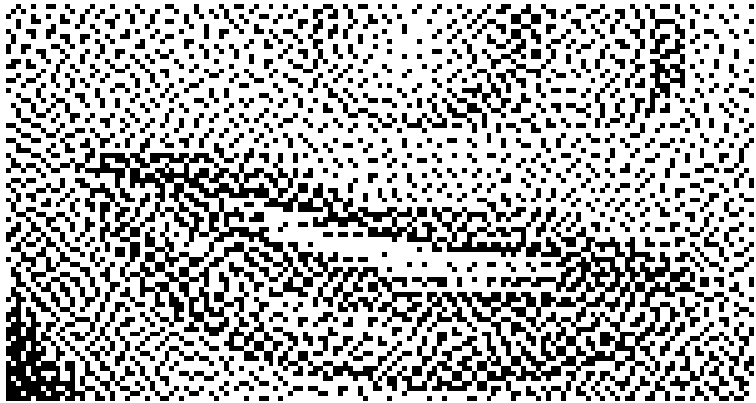}
\end{center}
\caption{\footnotesize 
A part (a local structure) of the resulting halftone image 
(the `mouth' of the woman shown in FIG. \ref{fig:fg02} (upper right)). 
Curious clusters of vortex configurations are observed. 
}
\label{fig:fg02_2}
\end{figure}

In FIG. \ref{fig:fg02}, we show the results. 
From the structure of the binary dots (see FIG. \ref{fig:fg02} (upper right)), 
we find that the resulting halftoned 
image looks well globally, however, locally it 
contains some curious clusters of vortex configurations 
(see FIG. \ref{fig:fg02_2}). 
The similar phenomena are generally observed 
in the halftoned binary dots image via 
the {\it error-diffusion method}.  
\mbox{} 

In the lower right panel of 
FIG. \ref{fig:fg02}, we show the recognized image $\mbox{\boldmath $s$}$. 
We find that the recognized image is quite similar to 
the original grayscale image and 
this result is a justification for our assumption to construct
the recognition model (\ref{eq:def_filter}). 
From these pictures, 
we might conclude that 
our algorithm works very well and 
the resulting halftoned binary dots 
look fine to represent the original grayscale levels. 
However, it is quite important for us to 
evaluate the performance quantitatively. 
In following , we evaluate 
the performance of the halftoning quantitatively. 

\begin{figure}[ht]
\begin{center}
\includegraphics[width=8.8cm]{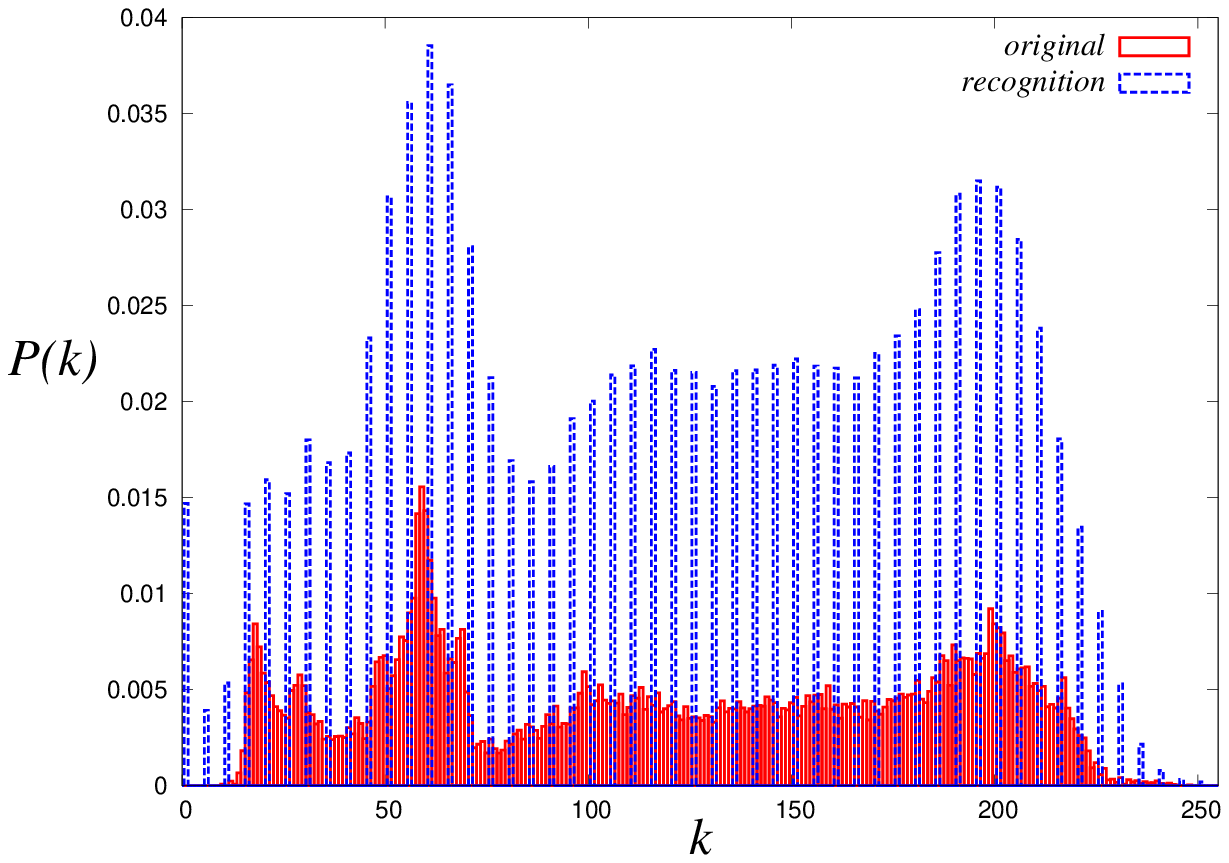}
\includegraphics[width=8.8cm]{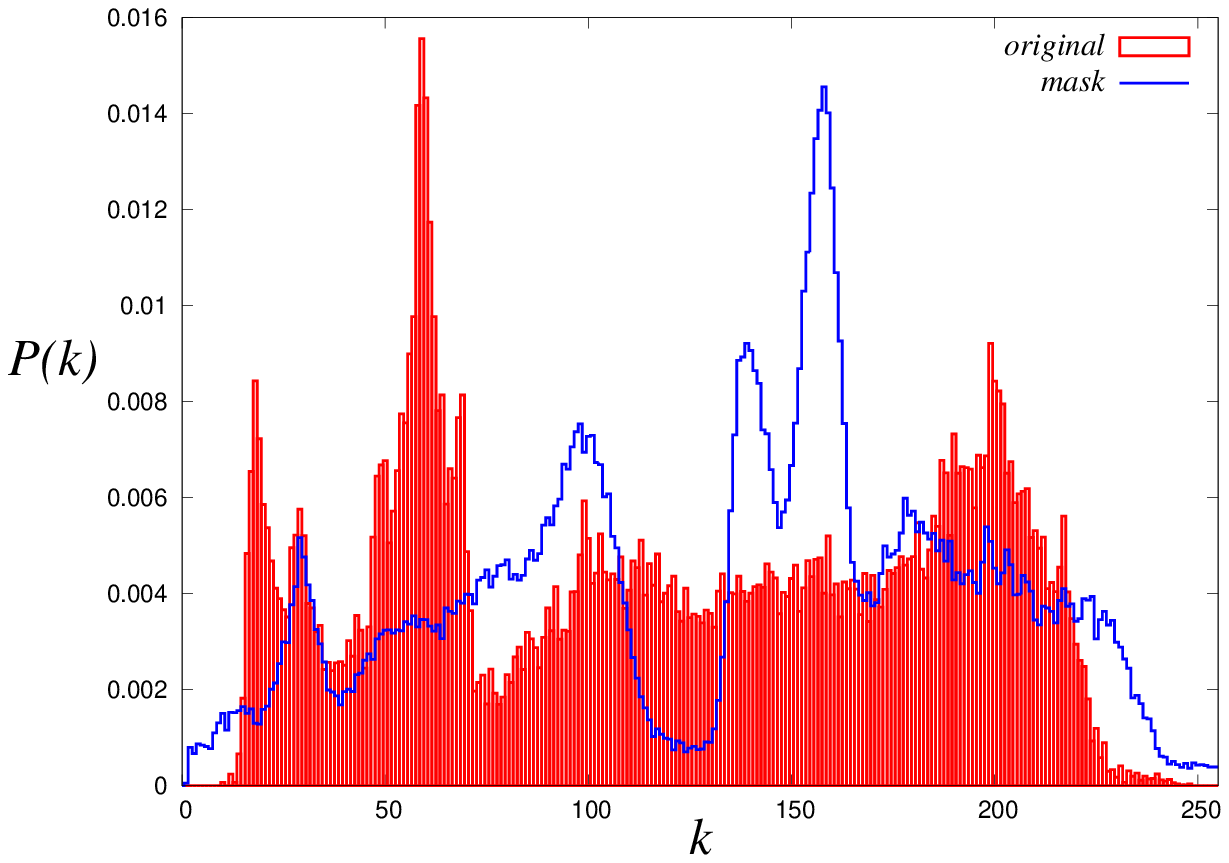}
\end{center}
\caption{\footnotesize 
Histograms (empirical distributions) $P(k)$ of 
the grayscale levels $k \in \{0,255\}$ for 
the 
original, the mask and the recognition images.  
}
\label{fig:fg02_3}
\end{figure}

In FIG. \ref{fig:fg02_3} (right), we show the histograms of 
grayscale levels for 
the original and recognition images.  
We find that the both shapes of the 
histograms have a similarity, however, 
there is a gap between the histograms in their height. 
Apparently, 
if the perfect minimization of the 
Hamiltonian is achieved, these two histogram should coincide 
with each other. Some theoretical argument on this issue is 
given in Appendix \ref{app:appA}. 

As we mentioned, 
this gap comes from the fact that 
the grayscale levels in the recognition image 
are restricted (defected) to 
$s_{x,y}=0,5,10,15,\cdots,255$ due to 
the definition of the weight $W_{x-i,y-j}$ with 
size $|\mathbb{N}(x,y)| \ll 256$. 
We discuss this issue later. 
In the right panel of FIG. \ref{fig:fg02_3}, we also show the histograms of 
grayscale levels for the original and threshold mask. 
Obviously, these two histograms are completely different as we also see the difference 
clearly from the upper left and the lower left panels in FIG. \ref{fig:fg02}.  

From the view point of 
human visual systems, 
we check the power-spectrum 
of the threshold mask and the halftone binary dots. 
In order to investigate the statistical 
properties of the model, we next investigate 
the power spectrum of the threshold mask and 
the halftone binary dots. 
Namely, 
we evaluate 
\begin{eqnarray}
P_{O}(f_{x},f_{y}) & = & 
\left\{
\frac{1}{L^{2}}
\sum_{i,j=1}^{L}
o_{i,j}
\cos \left[
\frac{2\pi}{L}
(if_{x}+jf_{y})
\right]
\right\}^{2} + 
\left\{
\frac{1}{L^{2}}
\sum_{i,j=1}^{L}
o_{i,j}
\sin \left[
\frac{2\pi}{L}
(if_{x}+jf_{y})
\right]
\right\}^{2}
\end{eqnarray}
where we should replace $o_{i,j}$ by 
the threshold mask 
$t_{i,j}$ or the 
halftoned binary dots 
$h_{i,j}$. 
We should bear in mind that here we consider the case 
$L_{1}=L_{2}=L$. 

\begin{figure}[ht]
\begin{center}
\mbox{}\hspace{-2cm}
\includegraphics[width=12cm]{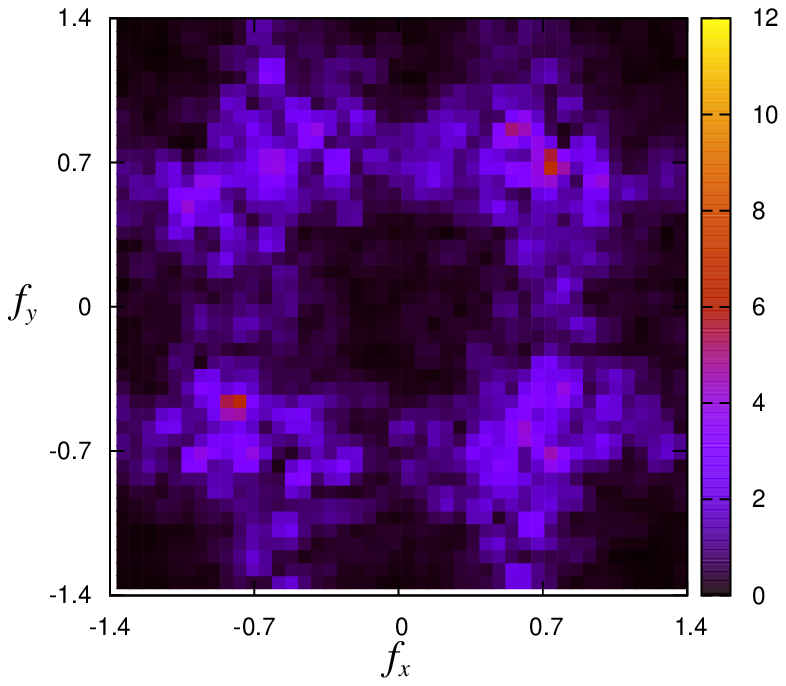} \hspace{-4.5cm}
\includegraphics[width=12cm]{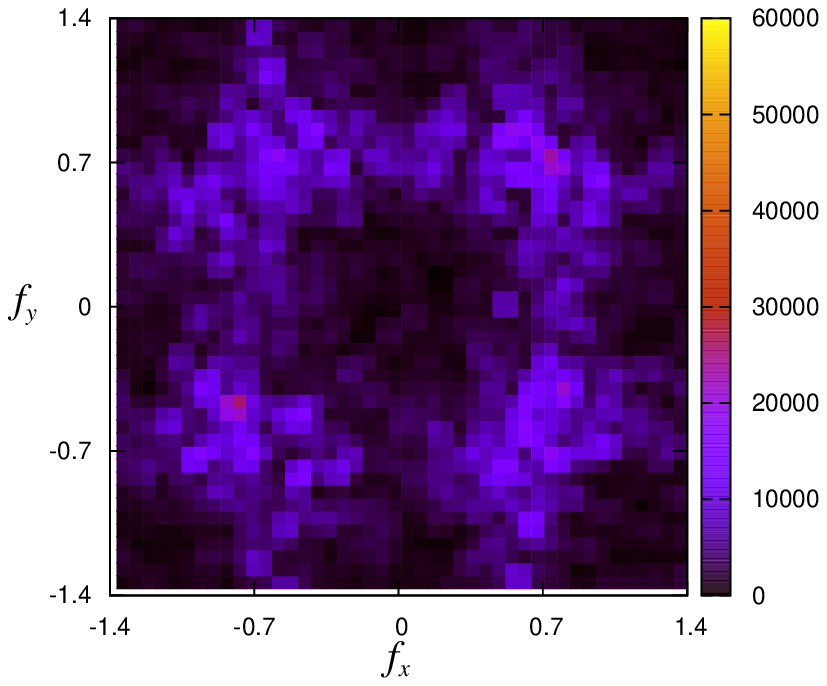}
\end{center}
\caption{\footnotesize 
The power-spectrum of 
the halftone image (left)  and threshold mask. 
}
\label{fig:fg_power_HM}
\end{figure}

\begin{figure}[ht]
\begin{center}
\includegraphics[width=8.8cm]{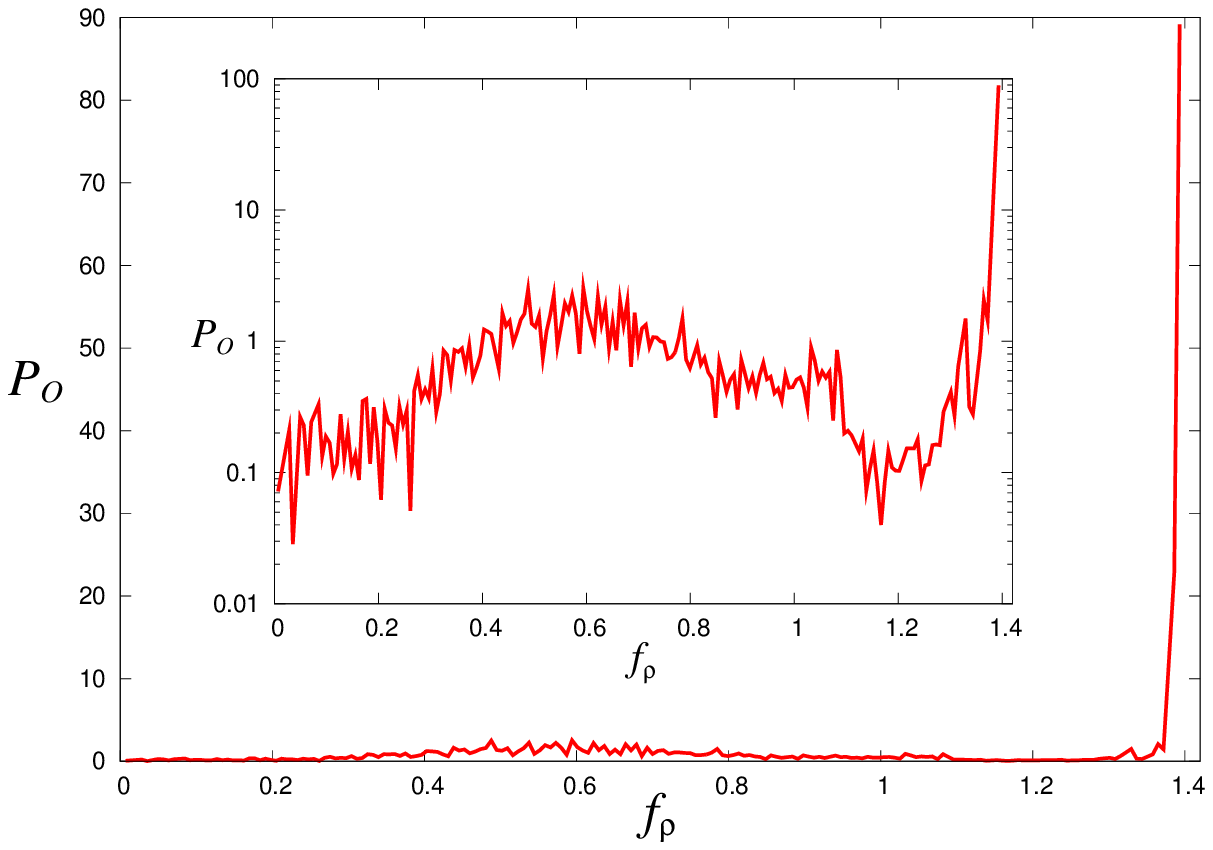}
\includegraphics[width=8.8cm]{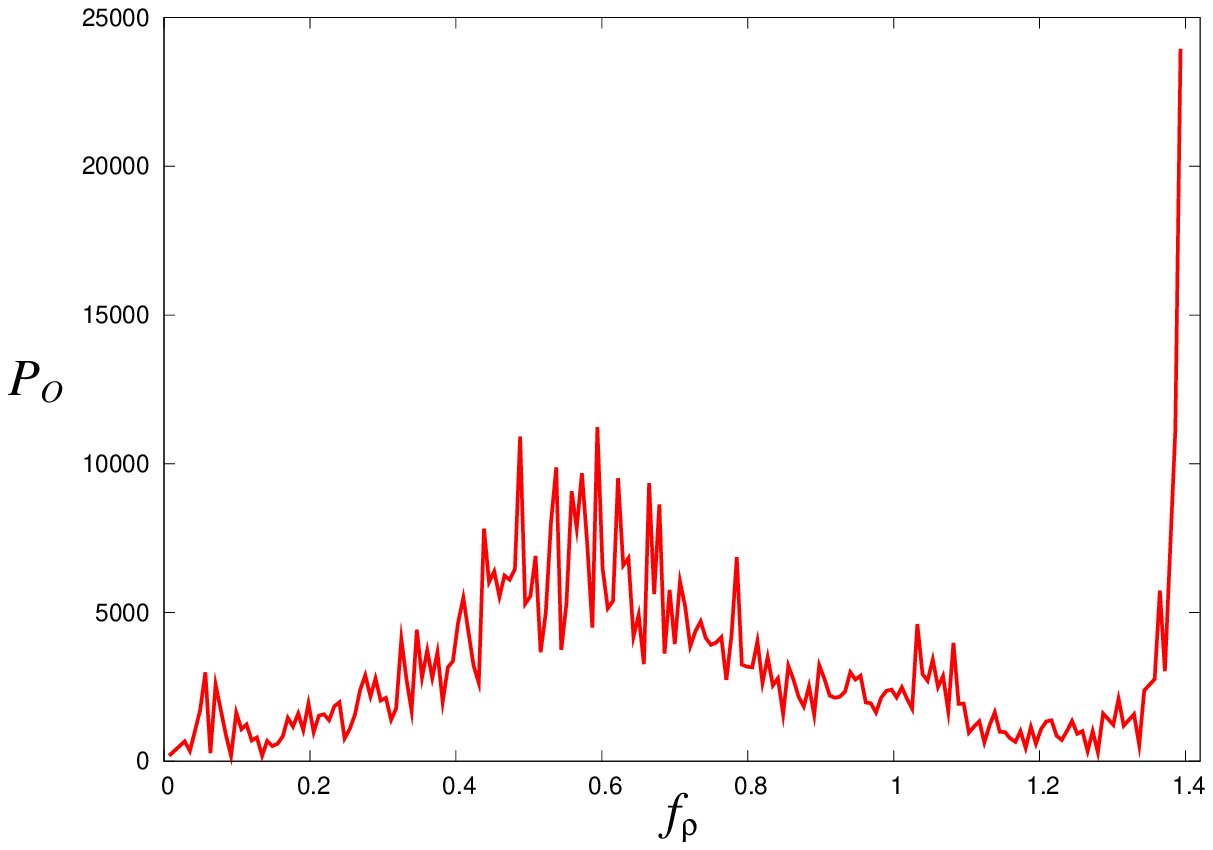}
\end{center}
\caption{\footnotesize 
The radial-averaged power spectrum 
of the halftoned image (left) and the threshold mask (right).  
The inset in the left panel is the same plot as in the panel in the logarithmic scale.
}
\label{fig:fg_power_r}
\end{figure}

We show the results in FIG. \ref{fig:fg_power_HM}. 
As well-known, the so-called blue noise mask, 
there exists principal frequency and below the value, 
the power spectrum drops to zero, whereas, 
the high-frequency components remains finite. 
On the other hand, the threshold mask generated 
by our algorithm apparently depends on 
the structure of the original grayscale image. 
Therefore, it is assumed that 
some of the spatial structures 
of the original image, especially, 
the smoothness leading up to 
the low frequency components 
in the power spectrum remains finite. 
However, from the power spectrum 
$P_{t}(f_{x},f_{y})$ shown in FIG.\ref{fig:fg_power_r}, 
we find that the low-frequency components almost 
disappear and high-frequency components remain finite. 

More convenient spectrum statistics 
is the so-called radial-averaged power 
spectrum \cite{Lau} defined by 
\begin{eqnarray}
P_{O}(f_{\rho}) & =& 
\frac{1}{N(\mathbb{R}(f_{\rho}))}
\sum_{f_{x},f_{y}
\in \mathbb{R}(f_{\rho})}
P_{O}(f_{x},f_{y}),
\end{eqnarray}
where 
$\mathbb{R}(f_{\rho})$ 
means the region of annular rings in the Fourier 
space with the radiuses $f_{\rho}$ and 
$f_{\rho}+\Delta_{\rho}$, 
namely, the 
width of the annular rings is given by $\Delta_{\rho}$. 
$N(\mathbb{R}(f_{\rho}))$ is the number of frequency samples in $R(f_{\rho})$. 
 
In FIG. \ref{fig:fg_power_r}, 
we plot the spectrum statistics for 
both threshold mask and resulting binary dots 
of the halftoned image. 
Inset of FIG. \ref{fig:fg_power_r} 
shows the same statistics 
for the threshold mask image. From this figure, 
we find that the low-frequency components are not 
dominant and relatively high-frequency 
components are larger than the low-frequency 
counter part. This tendency should be plausible  
from the view point of human-eyes modulation as 
we explained in FIG. \ref{fig:fg1}. 
\section{Inverse digital-halftoning}
\label{sec:INV}
In the previous sections, 
we consider the digital halftoning from the view point of statistical mechanics of spin systems. 
In engineering perspective, 
it is also important for us 
to retrieve the original grayscale image 
from a given 
halftoned binary dots 
when we attempt to capture 
the halftone image via scanner machines. 
The inverse problem of 
the digital halftoning is referred to as 
{\it inverse-halftoning}. 
In this section, we investigate 
the statistical performance 
of the inverse-halftoning 
by making use of the concept of statistical mechanics 
of disordered spin systems.  
Especially, 
we discuss the condition on which 
the Bayes-optimal inverse-halftoning is achieved.   
\subsection{Bayesian formula}
We first provide a Bayesian formula of inverse-halftoning. 
As we mentioned, the halftoned image $\mbox{\boldmath $h$}$ 
is given pixel-wise as $h_{x,y}=\Theta (g_{x,y}-t_{x,y}), \forall_{x,y}$. 
Thus, we might regard the process as a kind of `channel' which is 
written in terms of the following conditional probability: 
\begin{eqnarray}
P_{t}(\mbox{\boldmath $h$}|
\mbox{\boldmath $g$}) & = & 
\frac{\prod_{x,y}
\delta (h_{x,y}, 
\Theta (g_{x,y}-t_{x,y}))
}
{
\sum_{\mbox{\scriptsize \boldmath $h$}}
\prod_{x,y}
\delta (h_{x,y},
\Theta (g_{x,y}-t_{x,y}))
}
\end{eqnarray}
for a given threshold mask $\mbox{\boldmath $t$}$. 
Hence, the halftoning process is regarded as stationary memory-less 
deterministic erasure channel from the information theoretical point of view. 
When we choose the ferromagnetic prior, the posterior is obtained as 
\begin{eqnarray}
P(\mbox{\boldmath $\sigma$}|
\mbox{\boldmath $h$}: 
\mbox{\boldmath $t$}) & = & 
\frac{P_{t}(\mbox{\boldmath $h$}|\mbox{\boldmath $\sigma$})
P_{J}(\mbox{\boldmath $\sigma$})}
{\sum_{\mbox{\scriptsize \boldmath $\sigma$}}
P_{t}(\mbox{\boldmath $h$}|\mbox{\boldmath $\sigma$})
P_{J}(\mbox{\boldmath $\sigma$})} = 
\frac{{\exp}
[-J\sum_{\langle (x,y)(k,l) \rangle}
(\sigma_{x,y}-\sigma_{k,l})^{2}]
\prod_{x,y}
\delta (h_{x,y}-\Theta (g_{x,y}-t_{x,y}))}
{
\sum_{\mbox{\scriptsize \boldmath $\sigma$}}
{\exp}
[-J\sum_{\langle (x,y)(k,l) \rangle}
(\sigma_{x,y}-\sigma_{k,l})^{2}]
\prod_{x,y}
\delta (h_{x,y}-\Theta (g_{x,y}-t_{x,y}))}
\end{eqnarray}
where we defined 
$\sum_{\langle (x,y),(k,l) \rangle}(\cdots)$ 
as a sum over the nearest neighbouring pixel pairs.  
From the view point of statistical mechanics, 
the above posterior is regarded as the Boltzmann 
distribution of the Q-Ising model 
in which the mobility of each spin is tightly constrained in the range 
$t_{x,y} \leq g_{x,y} \leq Q-1$ for $h_{x,y}=1$ and 
$0  \leq g_{x,y} <t_{x,y}$ for $h_{x,y}=0$. 

For the above posterior, the Maximizer of Posterior Marginal 
(MPM for short) estimate is given by 
pixel-wise optimization of the posterior as  
\begin{eqnarray}
\hat{\sigma}_{x,y} & = & 
\arg\max_{\sigma_{x,y}}
P(\sigma_{x,y}|\mbox{\boldmath $h$}:
\mbox{\boldmath $t$}) \nonumber \\
\mbox{} & = & 
\Theta_{Q}
\left(
\frac{
\sum_{\mbox{\scriptsize \boldmath $\sigma$}}
\sigma_{x,y} \,
{\exp}
[-J\sum_{\langle (x,y)(k,l) \rangle}
(\sigma_{x,y}-\sigma_{k,l})^{2}]
\prod_{x,y}
\delta (h_{x,y}-\Theta (g_{x,y}-t_{x,y}))}
{
\sum_{\mbox{\scriptsize \boldmath $\sigma$}}
{\exp}
[-J\sum_{\langle (x,y)(k,l) \rangle}
(\sigma_{x,y}-\sigma_{k,l})^{2}]
\prod_{x,y}
\delta (h_{x,y}-\Theta (g_{x,y}-t_{x,y}))}
\right)
\end{eqnarray}
where $\Theta_{Q}(x)$ stands for a function 
to convert  a real value $x$ to 
the nearest integer.

In the inverse halftoning, we should solve $\forall_{x,y} \,\,\,
h_{x,y} = \Theta (g_{x,y}-t_{x,y})$ 
with respect to $\mbox{\boldmath $g$}$ for a given 
halftoned image $\mbox{\boldmath $h$}$ and 
the mask $\mbox{\boldmath $t$}$. 

In the reference \cite{Tadaki}, one of the authors considered restoration processes of 
grayscale images by making use of bit-decomposed data. 
Namely, 
we generate the $Q$ binary images whose pixel 
is given by 
$\forall_{x,y}\,\,\,h_{x,y}^{(m)}=
\Theta (g_{i}-m),\,
m \in \{0,\cdots,Q-1\}$ and 
transmitting them through some noise channels. 
Then, the paper \cite{Tadaki} dealt with 
the procedure to retrieve the original image 
$\mbox{\boldmath $g$}$ from the 
degraded $Q$ slices of the binary images 
\begin{eqnarray}
\mbox{\boldmath $h^{'}$}^{(m)} & = & 
\mbox{\boldmath $h$}^{(m)} + \mbox{\boldmath $\eta$}^{(m)} \,\,\,\,\,
(m=0,1,\cdots, Q-1)
\end{eqnarray}
where we defined 
$\mbox{\boldmath $h^{'}$}^{(m)} \equiv
\{h^{' (m)}_{x,y} |x=1,\cdots,L_{1}, 
y=1,\cdots, L_{2}\}$ and 
$\mbox{\boldmath $\eta$}^{(m)} \equiv
\{\eta_{x,y}^{(m)}|x=1,\cdots,L_{1}, 
y=1,\cdots, L_{2}\}$ stands for the additive noise. 
If there is no degrading process, 
it is obvious that the vector 
\begin{eqnarray}
\mbox{\boldmath $\hat{g}$} & \equiv  & 
\sum_{m=0}^{Q-1}
\mbox{\boldmath $h$}^{(m)}
\end{eqnarray}
is identical to the original grayscale image $\mbox{\boldmath $g$}$. 
However, in the present inverse-halftoning case, 
the only information we have is just a single slice $\mbox{\boldmath $h$}^{(m)}$. 
This fact makes the problem hard to treat. 

Obviously, this inverse-halftoning  is a typical ill-posed problem because 
there are a lot of candidates to satisfy the equations. 
In Appendix \ref{app:appB}, we evaluate two relevant quantities, namely, degree of degeneracy for 
possible solutions and 
mutual information to evaluate the difficulties of the problem.

In the next subsection, we discuss the relationship 
between the so-called Bayes-optimal solution and the 
Nishimori line established in the research field of spin glasses. 
\subsection{Bayes-optimal inverse-halftoning and the condition to achieve it}
In the previous studies \cite{Saika2009}, 
we investigated the inverse-halftoning 
on the bases of 
Markov chain Monte Carlo simulations and 
analysis of the infinite-range mean-field model. 
Then, we found several conditions 
on the hyper-parameters 
appearing in the Hamiltonian  
that gives a minimum of the mean square error 
numerically. 
However, so far, we do not yet provide 
any mathematically 
rigorous results on the performance of 
the inverse-halftoning 
defined on realistic two dimensional 
square lattices. 

In this section, we attempt to prove that 
the Bayes-optimal inverse-halftoning 
is achieved on a specific condition which is similar to the so-called Nishimori line. 
For the purpose, let us use here an alternative 
definition of pixel index. Namely, 
for $x,y=1,2,\cdots, L$,  we change the index by means of 
$i=x+Ly-L, i=1,2,\cdots,L^{2} \equiv N$ 
(conversely, 
$x=(i \mod L), y=[i/L]$). 
Then, we consider the 
true prior and the true inverse-halftoning process as follows. 
\begin{eqnarray}
P_{J_{0}}(\mbox{\boldmath $g$}) & = & 
\frac{{\exp}[-J_{0}\sum_{ij}
(g_{i}-g_{j})^{2}]
}
{
\sum_{\mbox{\scriptsize \boldmath $g$}}
{\exp}[-J_{0}\sum_{\langle ij \rangle}
(g_{i}-g_{j})^{2}]},\,\,\,\,
P_{t}(\mbox{\boldmath $h$}|
\mbox{\boldmath $g$}) 
= 
\frac{\prod_{i}
\delta(h_{i},\Theta(g_{i}-t_{i}))}
{
\sum_{\mbox{\scriptsize \boldmath $h$}}\prod_{i}
\delta(h_{i},\Theta(g_{i}-t_{i}))}
\label{eq:conventional}
\end{eqnarray}
where we defined 
the original image 
$\mbox{\boldmath $g$}=\{g_{i}=1,\cdots,Q|
i=1,\cdots,N\}$ and 
the halftoned image 
$\mbox{\boldmath $h$}=
\{h_{i}=0,1|i=0,\cdots,N\}$. 
We also used the definition 
$\langle ij \rangle$ to represent all 
nearest neighbouring pairs on the arbitrary 
lattice in finite dimension. 
The sums 
$\sum_{\mbox{\scriptsize \boldmath $g$}}(\cdots)$ and 
$\sum_{\mbox{\scriptsize \boldmath $h$}}(\cdots)$ 
denote 
$\sum_{\mbox{\scriptsize \boldmath $g$}}=
\prod_{i}
\sum_{g_{i}=1,\cdots,Q}(\cdots)$ and 
$\sum_{\mbox{\scriptsize \boldmath $h$}}=
\prod_{i} \sum_{h_{i}=0,1}
(\cdots)$, 
respectively. 
It should be noted that 
the above likelihood 
is another representation of 
the following dither method 
for each pixel 
\begin{eqnarray}
\forall_{i} \,\,\,\,
h_{i} & = &  
\Theta (g_{i}-t_{i}).
\end{eqnarray}
For this original 
image and 
the halftone process given by the likelihood 
(\ref{eq:conventional}), 
we naturally use the 
following 
posterior: 
\begin{eqnarray}
P_{J}(\mbox{\boldmath $\sigma$}|
\mbox{\boldmath $h$}) & = & 
\frac{
P_{t}(\mbox{\boldmath $h$}|
\mbox{\boldmath $\sigma$})
P_{J_{0}}(\mbox{\boldmath $\sigma$})}
{
\sum_{\mbox{\scriptsize \boldmath $\sigma$},
\mbox{\scriptsize \boldmath $h$}}
P_{t}(\mbox{\boldmath $h$}|
\mbox{\boldmath $\sigma$})
P_{J_{0}}(\mbox{\boldmath $\sigma$})
}
 = 
\frac{{\exp}[-J \sum_{\langle ij \rangle}
(\sigma_{i}-\sigma_{j})^{2}]
\prod_{i}
\delta(h_{i},\Theta(\sigma_{i}-t_{i}))
}
{
\sum_{\mbox{\scriptsize \boldmath $\sigma$},
\mbox{\scriptsize \boldmath $h$}}
{\exp}[-J \sum_{\langle ij \rangle}
(\sigma_{i}-\sigma_{j})^{2}]
\prod_{i}
\delta(h_{i},\Theta(\sigma_{i}-t_{i}))
}
\end{eqnarray}
The quantity to be evaluated is 
the following mean square error $D$ 
for an arbitrary pixel $i$: 
\begin{eqnarray}
\mbox{} & \mbox{} & 
D = \mathbb{E}_{
\mbox{\scriptsize \boldmath $g$},
\mbox{\scriptsize \boldmath $h$}}
[\{g_{i}-
\Theta_{Q}(\langle \sigma_{i} \rangle_{J})\}^{2}] = 
\frac{
\sum_{\mbox{\scriptsize \boldmath $g$},
\mbox{\scriptsize \boldmath $h$}}
{\exp}[-J_{0}\sum_{\langle ij \rangle}
(g_{i}-g_{j})^{2}]
\prod_{i}
\delta(h_{i},\Theta(g_{i}-t_{i}))
\left\{
g_{i}
-
\Theta_{Q}
\left(
\langle \sigma_{i} \rangle_{J} 
\right)
\right\}^{2}
}
{
\sum_{\mbox{\scriptsize \boldmath $g$},
\mbox{\scriptsize \boldmath $h$}}
{\exp}[-J_{0}\sum_{\langle ij \rangle}
(g_{i}-g_{j})^{2}]
\prod_{i}
\delta(h_{i},\Theta(g_{i}-t_{i}))
}
\end{eqnarray}
where 
$\mathbb{E}_{\mbox{\scriptsize 
\boldmath $g$},\mbox{\scriptsize \boldmath $h$}}[\cdots] 
\equiv \int \int 
P_{t}(\mbox{\boldmath $h$}|\mbox{\boldmath $g$})
P_{J_{0}}
(\mbox{\boldmath $g$})
(\cdots) d\mbox{\boldmath $g$}
d\mbox{\boldmath $h$}$ 
and we defined 
the average of a 
single pixel over 
the posterior by 
\begin{eqnarray}
\langle \sigma_{i} \rangle_{J} & \equiv & 
\frac{
\sum_{\mbox{\scriptsize \boldmath $\sigma$}}
\sigma_{i}
{\exp}[-J\sum_{\langle ij \rangle}
(\sigma_{i}-\sigma_{j})^{2}]
\prod_{i}
\delta(h_{i},\Theta(\sigma_{i}-t_{i}))
}
{
\sum_{\mbox{\scriptsize \boldmath $\sigma$},
\mbox{\scriptsize \boldmath $h$}}
{\exp}[-J\sum_{\langle ij \rangle}
(\sigma_{i}-\sigma_{j})^{2}]
\prod_{i}
\delta(h_{i},\Theta(\sigma_{i}-t_{i}))}.
\label{eq:proof_post}
\end{eqnarray}
To proceed the proof, 
we should notice that 
for any stochastic variable 
$A$, the following fact is satisfied
$\mathbb{E}_{A}[ \{A-\mathbb{E}_{A}[A]\}^{2}] = 
\mathbb{E}_{A}[A^{2}] - 
\mathbb{E}_{A}[A]^{2} \,\geq \,0$, 
namely, $\mathbb{E}_{A}[A^{2}] \,\geq \, \mathbb{E}_{A}[A]^{2}$. 
When we set 
$A = g_{i}-
\Theta_{Q}(\langle \sigma_{i} \rangle_{J})$, 
we have the following inequality: 
\begin{eqnarray}
D= \mathbb{E}_{\mbox{\scriptsize \boldmath $g$},\mbox{\scriptsize \boldmath $h$}}
[\{g_{i}-
\Theta_{Q}(\langle \sigma_{i} \rangle_{J})\}^{2}] & \,\,\,\geq\,\,\, & 
\left\{
\mathbb{E}_{\mbox{\scriptsize \boldmath $g$},\mbox{\scriptsize \boldmath $h$}}
[g_{i}]
-
\mathbb{E}_{\mbox{\scriptsize \boldmath $g$},\mbox{\scriptsize \boldmath $h$}}
[\Theta_{Q}(\langle \sigma_{i} \rangle_{J})]
\right\}^{2}
\label{eq:inequality}
\end{eqnarray}
In following, 
we shall evaluate 
the lower bound of the 
$D$, that is, 
the right most term of the above inequality. 
Among the staffs of the right most part in 
the above inequality, 
$\mathbb{E}_{\mbox{\scriptsize \boldmath $g$},
\mbox{\scriptsize \boldmath $h$}}[g_{i}]$ is 
easily evaluated as 
\begin{eqnarray}
\mathbb{E}_{\mbox{\scriptsize \boldmath $g$},
\mbox{\scriptsize \boldmath $h$}}[g_{i}] & = & 
\frac{
\sum_{\mbox{\scriptsize \boldmath $g$},
\mbox{\scriptsize \boldmath $h$}}
g_{i}\,
{\exp}[-J_{0}\sum_{\langle ij \rangle}
(g_{i}-g_{j})^{2}]
\prod_{i}
\delta(h_{i},\Theta(g_{i}-t_{i}))
}
{
\sum_{\mbox{\scriptsize \boldmath $g$},
\mbox{\scriptsize \boldmath $h$}}
{\exp}[-J_{0}\sum_{\langle ij \rangle}
(g_{i}-g_{j})^{2}]
\prod_{i}
\delta(h_{i},\Theta(g_{i}-t_{i}))
} \nonumber \\
\mbox{} & = & 
\frac{
\sum_{\mbox{\scriptsize \boldmath $g$}}
g_{i}\,
{\exp}[-J_{0}\sum_{\langle ij \rangle}
(g_{i}-g_{j})^{2}]
\sum_{\mbox{\scriptsize \boldmath $h$}}
\prod_{i}
\delta(h_{i},\Theta(g_{i}-t_{i}))
}
{
\sum_{\mbox{\scriptsize \boldmath $g$}}
{\exp}[-J_{0}\sum_{\langle ij \rangle}
(g_{i}-g_{j})^{2}]
\sum_{\mbox{\scriptsize \boldmath $h$}}
\prod_{i}
\delta(h_{i},\Theta(g_{i}-t_{i}))
} \nonumber \\
\mbox{} & = & 
\frac{
\sum_{\mbox{\scriptsize \boldmath $g$}}
g_{i}\,
{\exp}[-J_{0}\sum_{\langle ij \rangle }
(g_{i}-g_{j})^{2}]
}
{
\sum_{\mbox{\scriptsize \boldmath $g$}}
{\exp}[-J_{0}\sum_{\langle ij \rangle}
(g_{i}-g_{j})^{2}]
}  =  
\left(
\frac{Z_{J_{0}}^{(g>t)}}{Z_{J_{0}}}
\right)\, m_{i}^{(J_{0},g>t)} + 
\left(
\frac{Z_{J_{0}}^{(g<t)}}{Z_{J_{0}}}
\right)\, m_{i}^{(J_{0},g<t)} \equiv  m_{i}^{(J_{0})}
\end{eqnarray}
where we defined 
\begin{eqnarray}
m_{i}^{(J_{0},g>t)} & = & 
\frac{
\sum_{\mbox{\scriptsize \boldmath $g$} 
\in \{g_{i}|g_{i}>t_{i}\}}
g_{i}\, {\exp}[-J_{0}\sum_{\langle ij \rangle}
(g_{i}-g_{j})^{2}]
}
{
\sum_{\mbox{\scriptsize \boldmath $g$} 
\in \{g_{i}|g_{i}>t_{i}\}}{\exp}[-J_{0}\sum_{\langle ij \rangle}
(g_{i}-g_{j})^{2}]
} \\
m_{i}^{(J_{0},g<t)} & = & 
\frac{
\sum_{\mbox{\scriptsize \boldmath $g$} 
\in \{g_{i}|g_{i}<t_{i}\}}
g_{i} \,{\exp}[-J_{0}\sum_{\langle ij \rangle}
(g_{i}-g_{j})^{2}]
}
{
\sum_{\mbox{\scriptsize \boldmath $g$} 
\in \{g_{i}|g_{i}<t_{i}\}}{\exp}[-J_{0}\sum_{\langle ij \rangle}
(g_{i}-g_{j})^{2}]
}
\end{eqnarray}
and 
\begin{eqnarray}
Z_{J_{0}}^{(g>t)} & = & 
\sum_{\mbox{\scriptsize \boldmath $g$} 
\in \{g_{i}|g_{i}>t_{i}\}}{\exp}[-J_{0}\sum_{\langle ij \rangle}
(g_{i}-g_{j})^{2}] \\
Z_{J_{0}}^{(g<t)} & = & 
\sum_{\mbox{\scriptsize \boldmath $g$} 
\in \{g_{i}|g_{i}<t_{i}\}}{\exp}[-J_{0}\sum_{\langle ij \rangle}
(g_{i}-g_{j})^{2}] \\
Z_{J_{0}} & = & 
\sum_{\mbox{\scriptsize \boldmath $g$}}
{\exp}[-J_{0}\sum_{\langle ij \rangle}
(g_{i}-g_{j})^{2}]
\end{eqnarray}
Therefore, 
$\mathbb{E}_{\mbox{\scriptsize \boldmath $g$},
\mbox{\scriptsize \boldmath $h$}}[g_{i}]$ 
is identical to the local magnetization 
$m_{i}^{(J_{0})}$ 
of the pure ferromagnetic Q-Ising model 
having the interaction strength $J_{0}$. 

On the other hand, the 
lest of the term 
appearing 
in the right hand side of 
the equation 
(\ref{eq:inequality}) 
is written by 
\begin{eqnarray}
\mbox{} & \mbox{} & 
\mathbb{E}_{\mbox{\scriptsize \boldmath $g$},
\mbox{\scriptsize \boldmath $h$}}[\Theta_{Q}(\langle \sigma_{i} 
\rangle_{J})] \nonumber \\
\mbox{} & = & 
\frac{
\sum_{\mbox{\scriptsize \boldmath $g$},\mbox{\scriptsize \boldmath $h$}}
{\exp}[-J_{0}\sum_{\langle ij \rangle}
(g_{i}-g_{j})^{2}]
\prod_{i}
\delta(h_{i},\Theta(g_{i}-t_{i}))
\Theta_{Q}
\left(
\frac{
\mbox{$\sum$}_{\mbox{\scriptsize 
\boldmath $\sigma$}}
\sigma_{i}
{\exp}[-J\mbox{$\sum$}_{\langle ij \rangle}
(\sigma_{i}-\sigma_{j})^{2}]
\mbox{$\prod$}_{i}
\delta(h_{i},\Theta(\sigma_{i}-t_{i}))
}
{
\mbox{$\sum$}_{\mbox{\scriptsize \boldmath $\sigma$}}
{\exp}[-J\mbox{$\sum$}_{\langle ij \rangle}
(\sigma_{i}-\sigma_{j})^{2}]
\mbox{$\prod$}_{i}
\delta(h_{i},\Theta(\sigma_{i}-t_{i}))
}
\right)
}
{
\mbox{$\sum$}_{\mbox{\scriptsize \boldmath $g$},
\mbox{\scriptsize \boldmath $h$}}
{\exp}[-J_{0}\mbox{$\sum$}_{\langle ij \rangle}
(g_{i}-g_{j})^{2}]
\prod_{i}
\delta(h_{i},\Theta(g_{i}-t_{i}))
} \nonumber \\
\mbox{} & = &  
\frac{
\sum_{\mbox{\scriptsize \boldmath $g$}}
{\exp}[-J_{0}\sum_{\langle ij \rangle}
(g_{i}-g_{j})^{2}]
\Theta_{Q}
\left(
\frac{
\mbox{$\sum$}_{\mbox{\scriptsize \boldmath $\sigma$}}
\sigma_{i}
{\exp}[-J\mbox{$\sum$}_{\langle ij \rangle}
(\sigma_{i}-\sigma_{j})^{2}]
\mbox{$\prod$}_{i}
\delta(\Theta(g_{i}-t_{i}),\Theta(\sigma_{i}-t_{i}))
}
{
\mbox{$\sum$}_{\mbox{\scriptsize \boldmath $\sigma$}}
{\exp}[-J\mbox{$\sum$}_{\langle ij \rangle}
(\sigma_{i}-\sigma_{j})^{2}]
\mbox{$\prod$}_{i}
\delta(\Theta(g_{i}-t_{i}),\Theta(\sigma_{i}-t_{i}))
}
\right)
}
{
\sum_{\mbox{\scriptsize \boldmath $g$}}
{\exp}[-J_{0}\sum_{\langle ij \rangle}
(g_{i}-g_{j})^{2}]
} \nonumber \\
\mbox{} & = & 
\left(
\frac{Z_{J_{0}}^{(g>t)}}{Z_{J_{0}}}
\right)
\Theta_{Q}
\left(
\frac{
\mbox{$\sum$}_{
\mbox{\scriptsize \boldmath $\sigma$} \in 
\{\sigma_{i}|\sigma_{i}>t_{i}\}}
\sigma_{i}
{\exp}[-J\mbox{$\sum$}_{\langle ij \rangle }
(\sigma_{i}-\sigma_{j})^{2}]
}
{
\mbox{$\sum$}_{\mbox
{\scriptsize \boldmath $\sigma$}
\{\sigma_{i}|\sigma_{i} >t_{i}\}}
{\exp}[-J\mbox{$\sum$}_{\langle ij \rangle}
(\sigma_{i}-\sigma_{j})^{2}]
}
\right)  \nonumber \\
\mbox{} & + & 
\left(
\frac{Z_{J_{0}}^{(g<t)}}{Z_{J_{0}}}
\right)
\Theta_{Q}
\left(
\frac{
\mbox{$\sum$}_{
\mbox{\scriptsize \boldmath $\sigma$} \in 
\{\sigma_{i}|\sigma_{i}<t_{i}\}}
\sigma_{i}
{\exp}[-J\mbox{$\sum$}_{\langle ij \rangle}
(\sigma_{i}-\sigma_{j})^{2}]
}
{
\mbox{$\sum$}_{\mbox
{\scriptsize \boldmath $\sigma$}
\{\sigma_{i}|\sigma_{i} <t_{i}\}}
{\exp}[-J\mbox{$\sum$}_{\langle ij \rangle}
(\sigma_{i}-\sigma_{j})^{2}]
}
\right) \nonumber \\
\mbox{} & = & 
\left(
\frac{Z_{J_{0}}^{(g>t)}}{Z_{J_{0}}}
\right) 
\Theta_{Q}(m_{i}^{(J,\sigma>t)}) + 
\left(
\frac{Z_{J_{0}}^{(g<t)}}{Z_{J_{0}}}
\right) 
\Theta_{Q}(m_{i}^{(J,\sigma<t)})
\end{eqnarray}
Therefore,  
the lower 
bound of the 
mean square error 
is evaluated as follows. 
\begin{eqnarray}
D & \geq & 
\{
\mathbb{E}_{\mbox{\scriptsize \boldmath $g$},
\mbox{\scriptsize \boldmath $h$}}[g_{i}] -
\mathbb{E}_{\mbox{\scriptsize \boldmath $g$},
\mbox{\scriptsize \boldmath $h$}}
[\Theta_{Q}(\langle \sigma_{i} \rangle_{J})] 
\}^{2} \nonumber \\
\mbox{} & = & 
\left\{
\left(
\frac{Z_{J_{0}}^{(g>t)}}{Z_{J_{0}}}
\right)
\left(
m_{i}^{J_{0},g>t}-
\Theta_{Q}(m_{i}^{J,\sigma>t})
\right)
+
\left(
\frac{Z_{J_{0}}^{(g<t)}}{Z_{J_{0}}}
\right)
\left(
m_{i}^{J_{0},g<t}-
\Theta_{Q}(m_{i}^{J,\sigma<t})
\right)
\right\}^{2} \nonumber \\
\mbox{} & = & 
\left\{
\left(
\frac{Z_{J_{0}}^{(g>t)}}{Z_{J_{0}}}
\right)
\left|
m_{i}^{J_{0},g>t}-
\Theta_{Q}(m_{i}^{J,\sigma>t})
\right|
+
\left(
\frac{Z_{J_{0}}^{(g<t)}}{Z_{J_{0}}}
\right)
\left|
m_{i}^{J_{0},g<t}-
\Theta_{Q}(m_{i}^{J,\sigma<t})
\right|
\right\}^{2} \nonumber \\
\mbox{} & \geq & 
\left\{ 
\left(
\frac{Z_{J_{0}}^{(g>t)}}{Z_{J_{0}}}
\right) \delta_{1} + 
\left(
\frac{Z_{J_{0}}^{(g<t)}}{Z_{J_{0}}}
\right) \delta_{2}
\right\}^{2}
\end{eqnarray}
where 
the equality on the last line is 
satisfied for 
$J=J_{0}$. 
We also used the fact that 
the sign of 
$(
m_{i}^{J_{0},g>t}-
\Theta_{Q}(m_{i}^{J,\sigma<t}))$ and 
$(
m_{i}^{J_{0},g>t}-
\Theta_{Q}(m_{i}^{J,\sigma<t}))$ are the same 
because 
the local magnetizations 
$m_{i}^{J,\sigma>t}$ and 
$m_{i}^{J,\sigma<t}$ are monotonically  
increasing 
function with respect to $J$. 
$\delta=\delta_{1},\delta_{2}$ is a {\it quantization error} 
due to the Q-generalized 
step function 
$\Theta_{Q}(\cdots)$ defined as 
\begin{eqnarray}
\Theta_{Q}(x) & = & x -\delta 
\end{eqnarray}
From the argument we presented above, 
we found that 
the performance 
of the inverse-halftoning 
achieved 
by the posterior (\ref{eq:proof_post}) is 
optimized on the specific condition $J=J_{0}$ 
which is similar to the Nishimori line (point) \cite{Nishi1981} in the research field of spin glasses. 
\section{Simultaneous generation of mask, halftone and recognition images}
\label{sec:simult}
Finally we show that 
both halftoning and the inverse-halftoning processes are unified under a single Hamiltonian, namely, 
it is possible for us to obtain the threshold mask, 
the halftoned and inverse-halftoned  images simultaneously by finding 
the ground state of the spin systems. 

In this paper, we proposed the digital halftoning and the inverse-halftoning separately. 
However, from the form of the Hamiltonian 
$\mathcal{H}(\mbox{\boldmath $t$}|\mbox{\boldmath $g$})$, 
the definition of the halftoned image:  $\forall_{x,y}\,\,\,
h_{x,y}=\Theta (g_{x,y}-t_{x,y})$ and 
the recognition image:  $\forall_{x,y}\,\,\,
s_{x,y}=\sum_{i,j\in \mathbb{N}(x,y)}
W_{x-i,y-j}\Theta (g_{x,y}-t_{x,y})$, 
it is confirmed that these three important 
images are obtained simultaneously 
in the single theoretical framework, 
that is, minimizing the Hamiltonian 
$\cal{H}(\mbox{\boldmath $t$}|\mbox{\boldmath $g$})$ 
with respect to the mask $\mbox{\boldmath $t$}$ 
for a given original image $\mbox{\boldmath $g$}$. 

Then, we should remember that 
there is a gap between 
the histograms of the grayscale levels for 
the recognition and the original images. 
This gap might make us hard to accept 
the recognition image as the inverse-halftoned one as a solution. 

As we mentioned before, 
this gap comes from the fact that 
the grayscale levels in the recognition image 
are restricted (defected) to 
$s_{x,y}=0,5,10,15,\cdots,255$ due to 
the definition of the weight $W_{x-i,y-j}$ with 
size $|\mathbb{N}(x,y)|=5 \times 5=25  \ll 256=Q$. 
To reduce the gap, we might use 
the following linear filter: 
\begin{eqnarray}
\forall_{x,y}\, \, \, \, \, s_{x,y}^{(t)} & = & 
\frac{1}{|\mathbb{N}^{'}(x,y)|} \sum_{k,l\in \mathbb{N}^{'}(x,y)}
s_{k,l}^{(t-1)}, \,\,\,t=1,\cdots,n
\label{eq:meanF}
\end{eqnarray}
where the initial condition $\mbox{\boldmath $s$}^{(0)}$ are chosen as 
$\forall_{x,y}\, \,\, s_{x,y}=\sum_{i,j \in \mathbb{N}(x,y)}W_{x-i,y-j}\Theta (g_{x,y}-t_{x,y})$ for 
the minimum energy state of the Hamiltonian 
$\mathcal{H}(\mbox{\boldmath $t$}|\mbox{\boldmath $g$})$. 
Thus, for a given $\mbox{\boldmath $s$}^{(0)}$, 
we recursively operate the above map (\ref{eq:meanF}) $n$ times,  and then, 
one might obtain 
more plausible image than the $\forall_{x,y}\,\,\, s_{x,y}^{(0)}$ 
as the inverse-halftoned image. 

\begin{figure}[ht]
\begin{center}
\includegraphics[width=8.8cm]{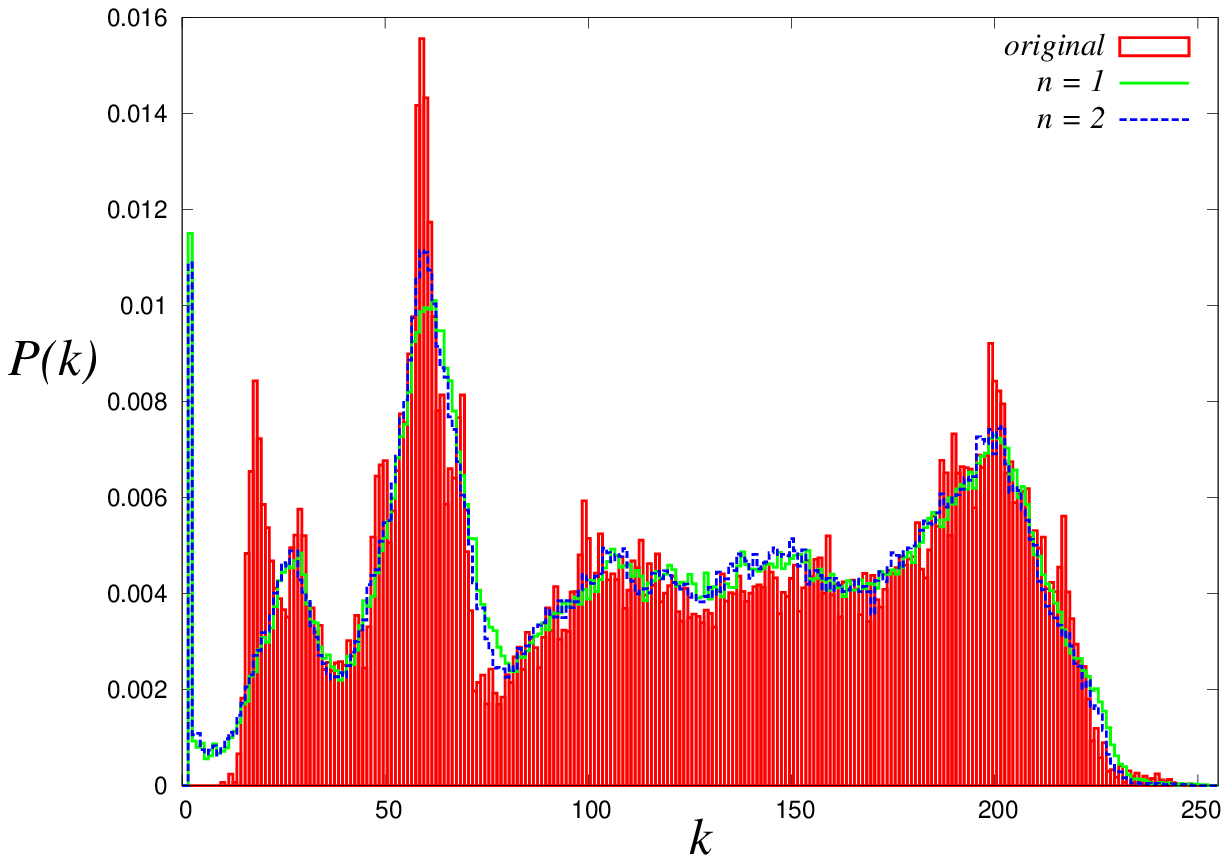}
\includegraphics[width=8.8cm]{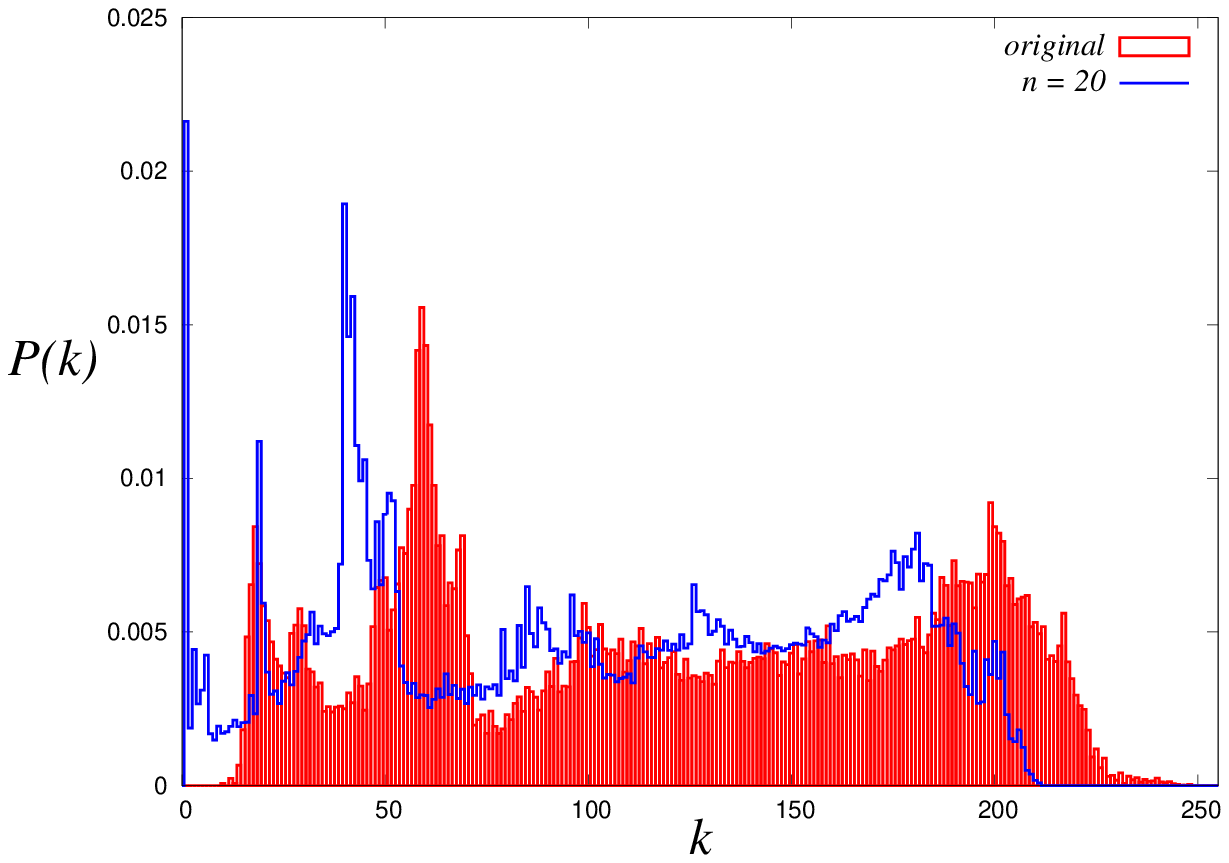}
\end{center}
\caption{\footnotesize 
Histogram of grayscale levels for 
the original image and the recognition images after operating 
the linear filter (\ref{eq:meanF}) with $n=1, 2$ (left) and $n=20$ (right). 
}
\label{fig:fg_hist_meanF}
\end{figure}

\begin{figure}[ht]
\begin{center}
\includegraphics[width=8.5cm]{origQ256_wmnG400B.eps} 
\includegraphics[width=8.5cm]{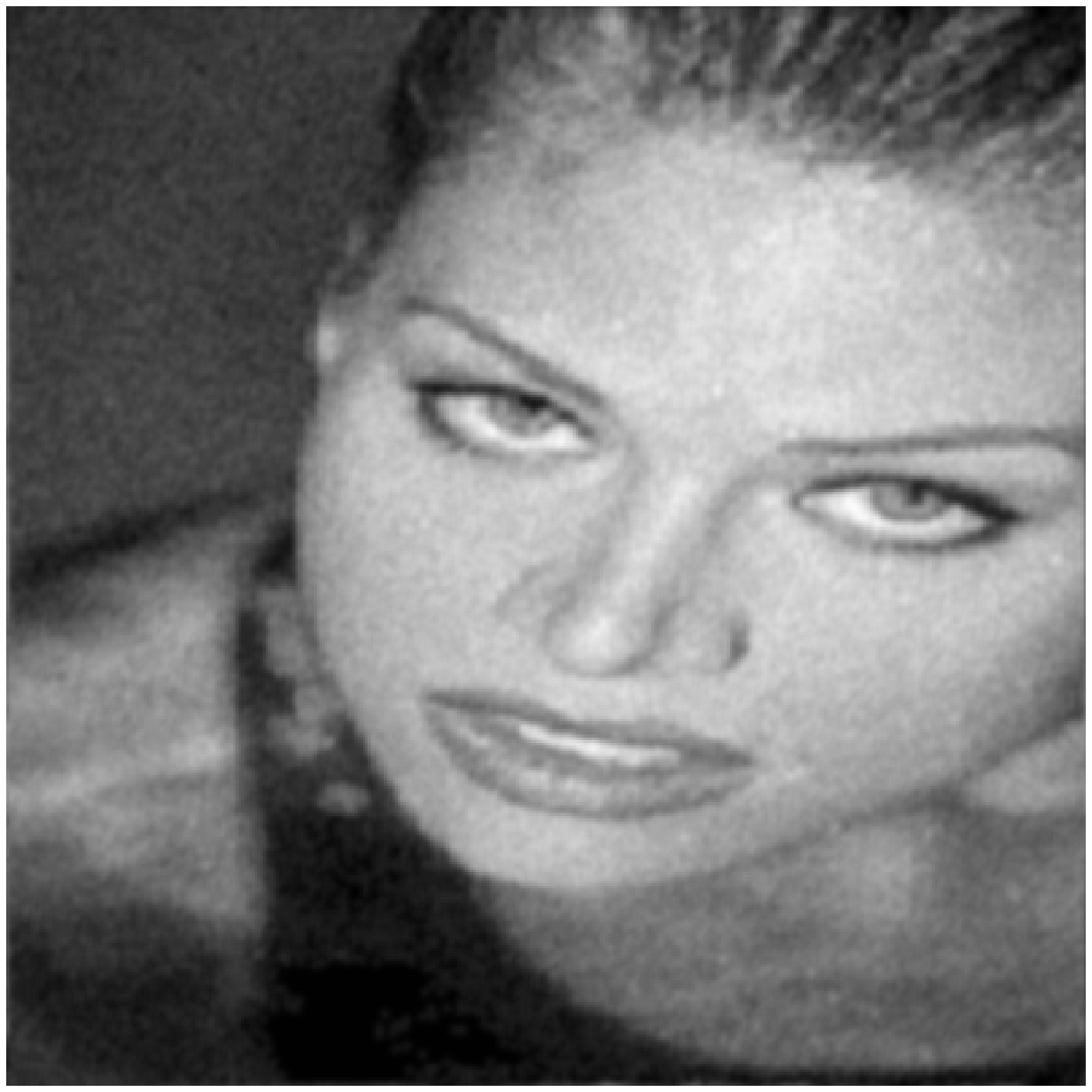} \\
\includegraphics[width=8.5cm]{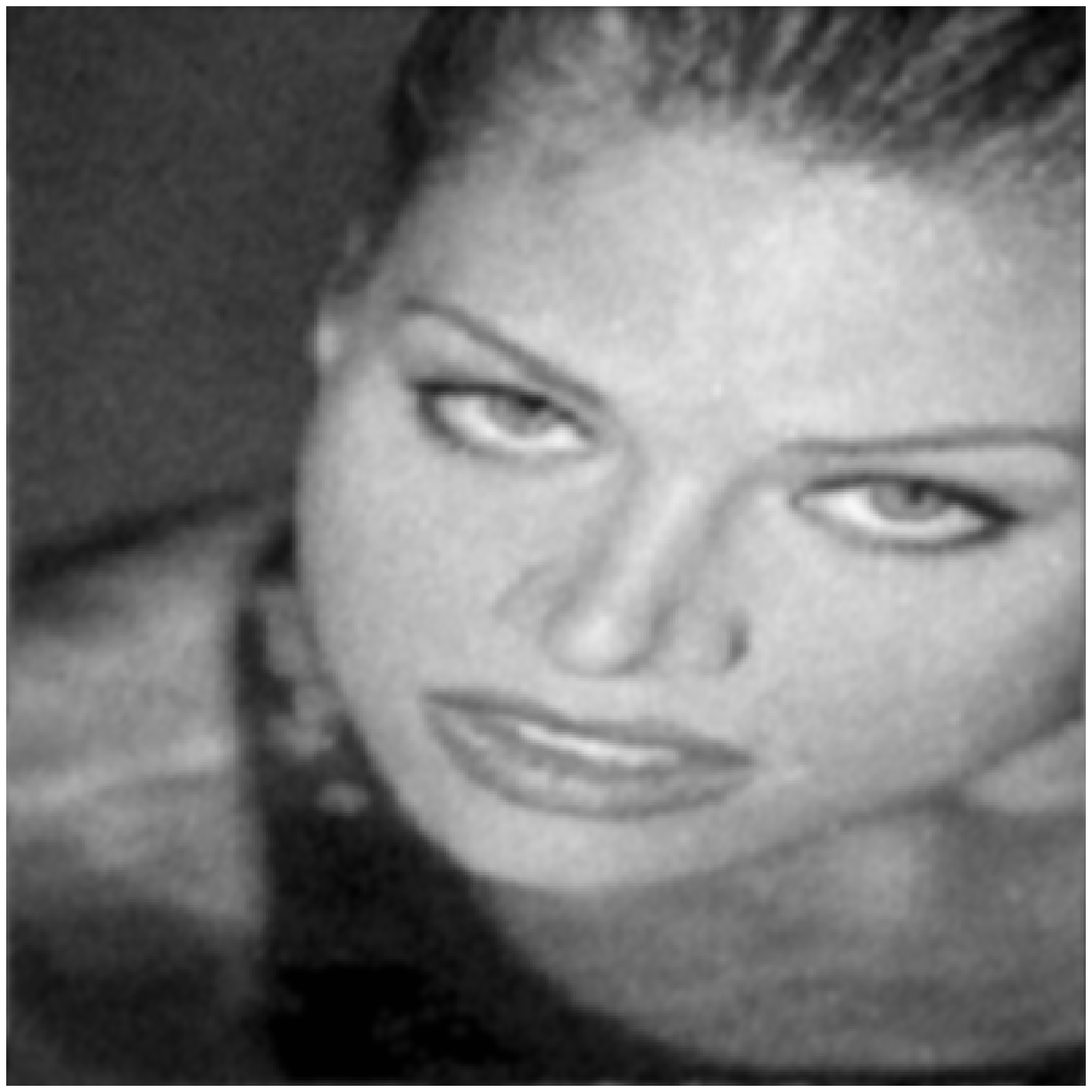} 
\includegraphics[width=8.5cm]{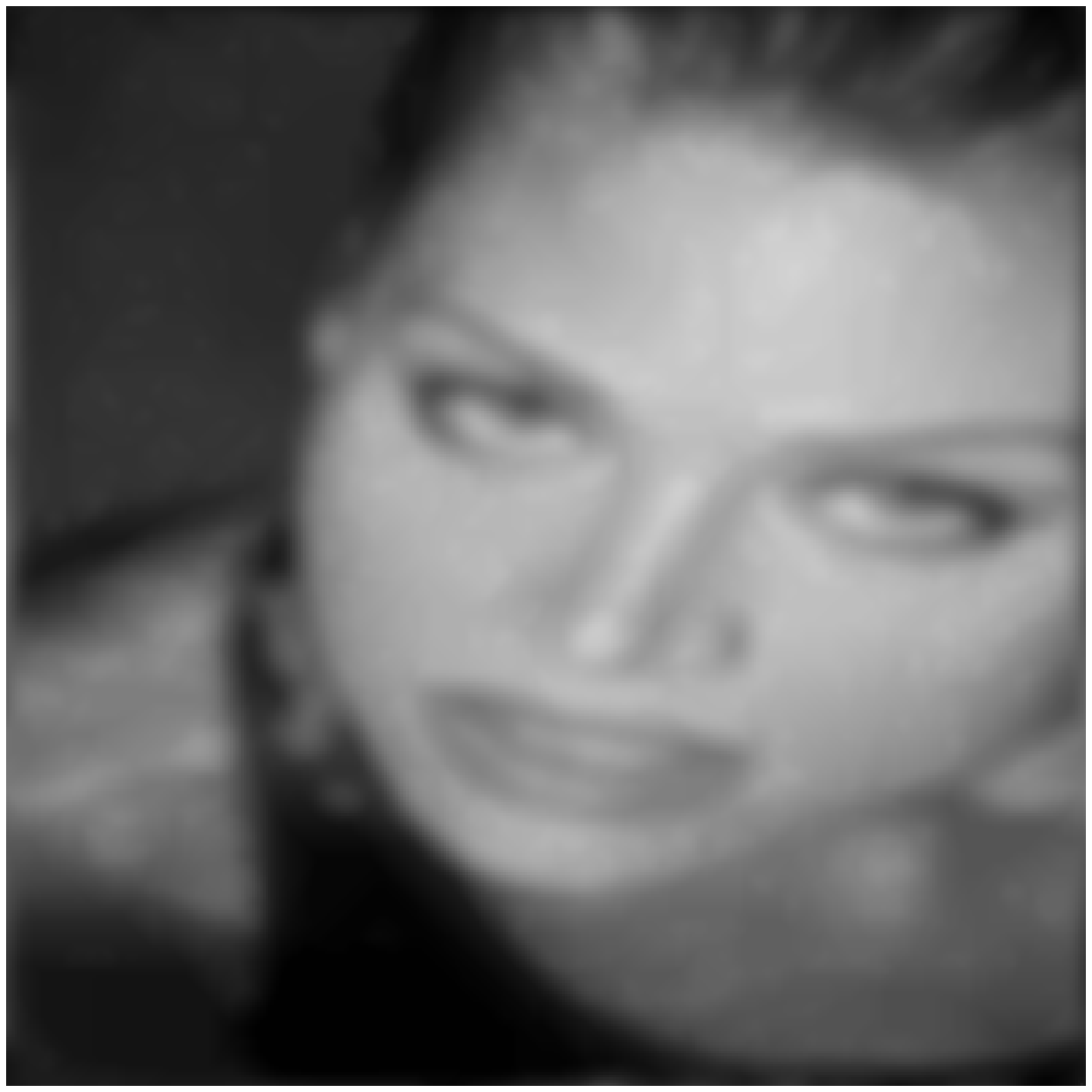} 
\end{center}
\caption{\footnotesize 
From the upper left to the lower right, 
the original image, the recognized images after operating 
the linear filter (\ref{eq:meanF}) with $n=1,2$ and $n=20$ are shown. 
}
\label{fig:fg_meanF}
\end{figure}

We show the resulting histogram of the grayscale levels 
in FIG. \ref{fig:fg_hist_meanF} and 
the corrected recognition images in 
FIG. \ref{fig:fg_meanF} for $n=1$ and $2$.  
We set the number of nearest neighbouring pixels around 
the pixel at $(x,y)$ as $|\mathbb{N}^{'}(x,y)|=3\times 3=9$, 
namely, 
$\mathbb{N}^{'}(x,y)=
\{
(x-1,y+1),(x,y+1),(x+1,y+1),
(x-1,y),(x,y),(x+1,y),
(x-1,y-1),(x,y-1),(x+1,y-1)\}$. 
From these figures, we find that the gap between two images actually reduced and 
the resulting recognition image is improved by `mixing effect' on 
the defected grayscale levels $s_{x,y}=0,5,10,15,\cdots,255$. 
However, when we increase the number of 
iteration $n$, the mixing effect works too much on the 
recognition image and it makes the local structure of the 
image too smooth (see FIG. \ref{fig:fg_hist_meanF} (right) and 
FIG. \ref{fig:fg_meanF} (lower right)).  
\section{Summary}
\label{sec:summary}
In this paper, we proposed a possible statistical-mechanical modeling for digital halftoning. 
This formalism helps us to understand 
the problem as a combinatorial optimization 
which is described by a sort of disordered spin systems. 
Finding the ground state was achieved by simulated annealing 
and we found that the resulting binary dots looks well 
to represent the original grayscale levels. 
The quality of the binary dots was evaluated from 
the power-spectrum statistics. 
We found that the binary dots contain relatively high frequency 
components which are plausible from the view point of human-eyes modulation properties.  
We also proposed a theoretical framework to 
evaluate the statistical performance of 
the inverse digital-halftoning based on statistical mechanics. 
From the Bayesian inference view point, we rigorously show that the 
Bayes-optimal inverse-halftoning 
is achieved on a specific condition which is very similar to the so-called 
Nishimori line in the research field of spin glasses. 
We hope our formulation might be applied to 
generating of binary halftoned images effectively and 
evaluating the performance for the inverse-halftoning from 
halftoned binary images obtained by various algorithms. 
\\

This work was financially supported by Grant-in-Aid, 
Scientific Research on Priority Areas Deepening 
and Expansion of Statistical Mechanical 
Informatics (DEX-SMI) of the Ministry of Education, 
Culture, Sports, Science and Technology (MEXT) No. 18079001. 
One of the authors (JI) was financially supported by 
Grant-in-Aid for Scientific Research (C) 
of Japan Society for 
the Promotion of Science, No. 22500195 and {\it INSA (Indian National Science Academy) -  JSPS 
(Japan Society of Promotion of Science)  Bilateral Exchange Programme}. 
\appendix
\section{Distribution of the recognized pixels}
\label{app:appA}
\begin{figure}[ht]
\begin{center}
\rotatebox{-90}{\includegraphics[width=5.8cm]{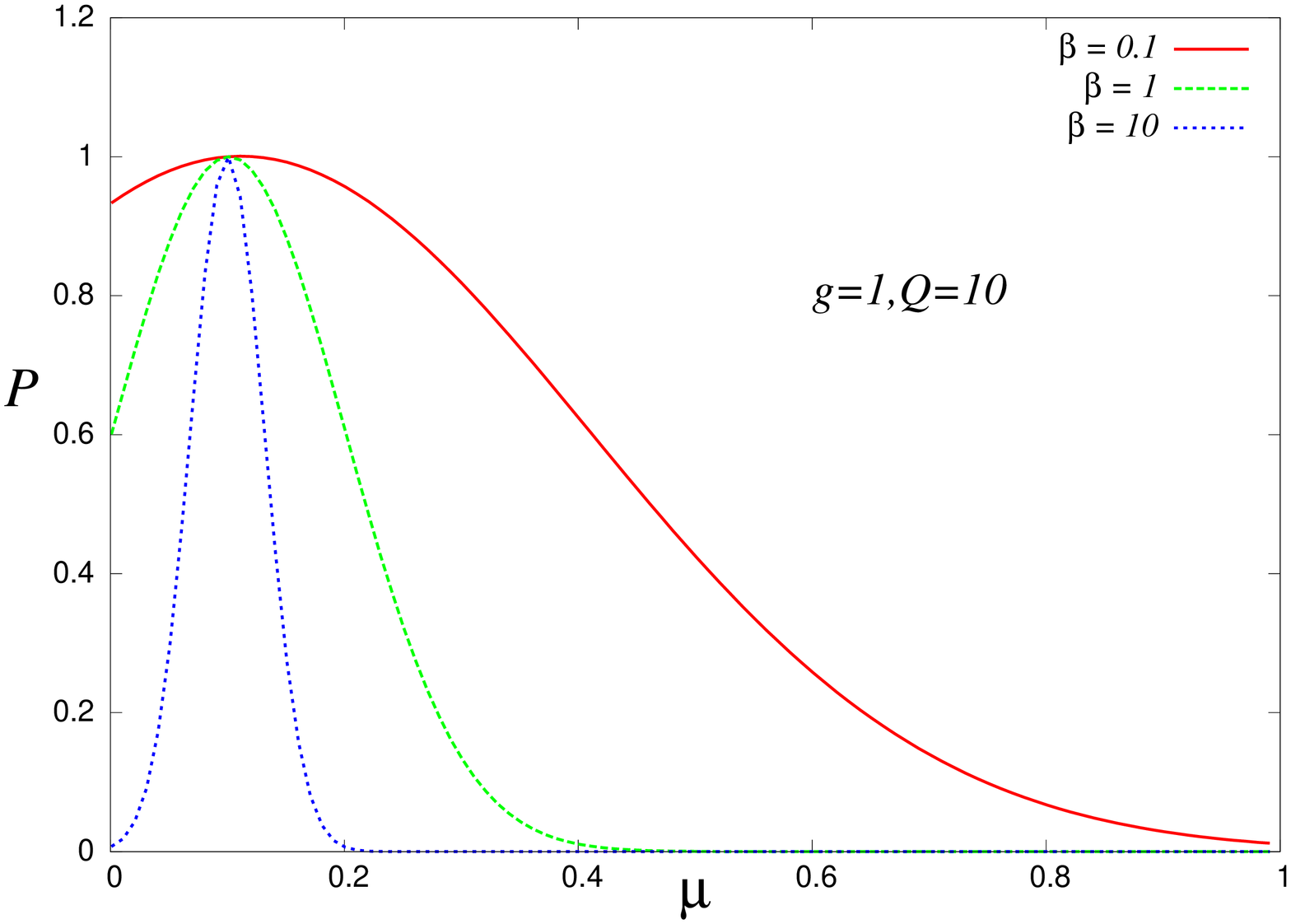}}
\rotatebox{-90}{\includegraphics[width=5.8cm]{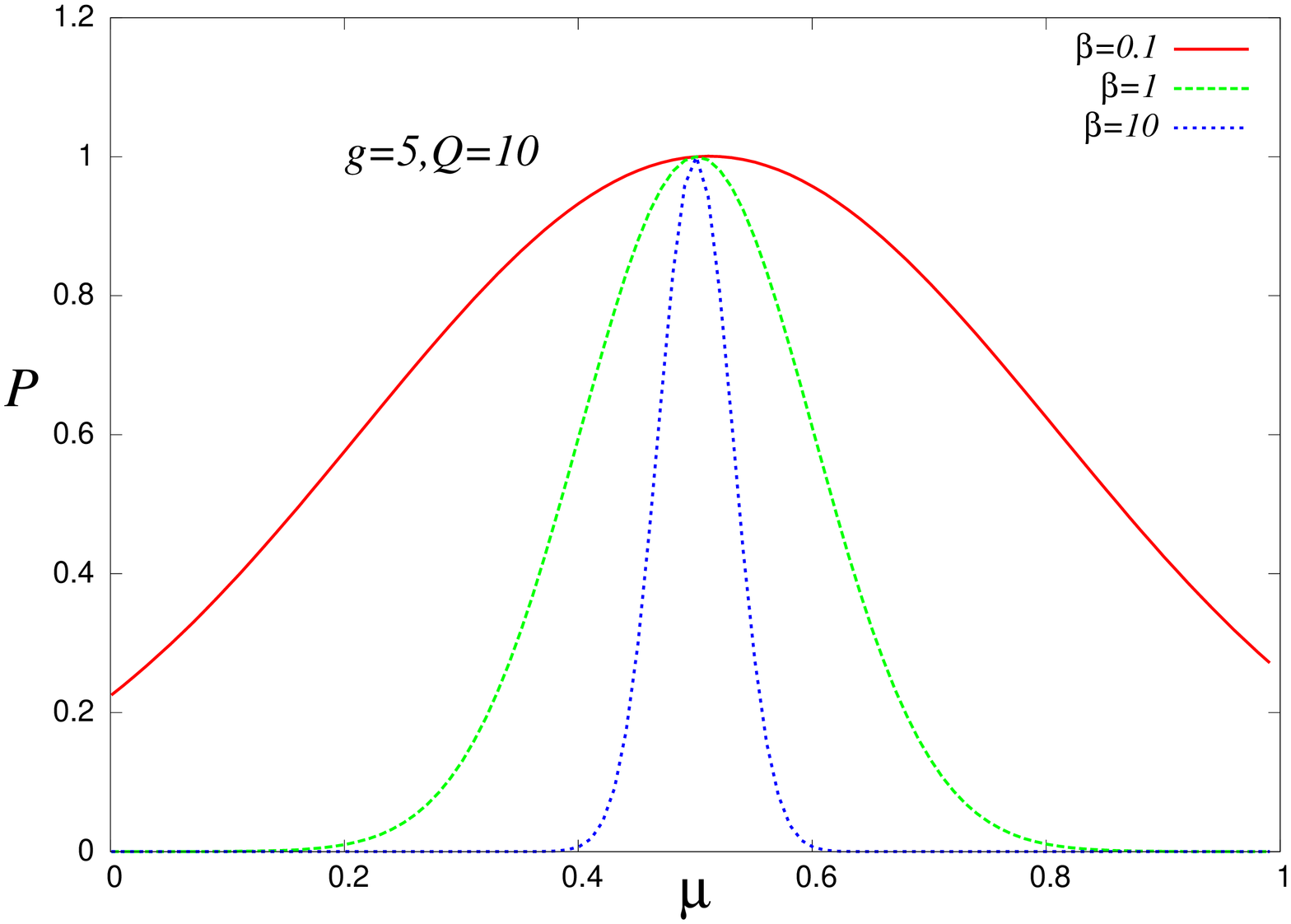}}
\end{center}
\caption{\footnotesize 
The distribution 
$P_{\mu,Q}(\mu)$ for several values of 
$Q$ and $\beta$ and $g$. 
}
\label{fig:fg_density}
\end{figure}

In Sec. \ref{sec:Numerical}, 
we evaluated the performance 
of halftoning 
through two different measurements, 
namely, the histograms of grayscale levels 
and the power-spectrum 
via computer simulations. However, 
it might be helpful for us to 
evaluate the performance analytically. 
In this Appendix, we derive the distribution of 
the recognized pixels, 
that is, 
the number of black pixels in a window 
with $|\mathbb{N}(x,y)|=Q (\ll 256)$ analytically. 
Namely, we consider the distribution of the 
following quantities: 
\begin{eqnarray}
\mu_{i} & = & 
\frac{1}{Q} 
\sum_{l \in \mathbb{N}(i)}
\Theta (g_{l}-t_{l}) = s_{i} 
\end{eqnarray}
where we consider the case of $L_{1}=L_{2}=L$ and defined the index 
$i$ so as to satisfy $i=x+Ly-L$, 
namely, $i$ takes 
$i=1,2,\cdots, L^{2}$ and 
we set $L^{2}=N$. 
It is obvious that 
$(x,y)$ for a given $i$ are given as 
$x=(i \mod L), y=[i/L]$. 
Here we consider the case of 
$|\mathbb{N}(i)|=Q=\mathcal{O}(1)$. 
Then, we obtain 
the general formula for the distribution as follows. 
\begin{eqnarray}
P_{\beta,Q}(\mu_{i}:\mbox{\boldmath $g$}) = 
\frac{
\prod_{i} \sum_{t_{i}=0}^{Q-1} 
\delta (Q\mu_{i} - 
\sum_{l \in \mathbb{N}(i)}
\Theta (g_{l}-t_{l})
) 
\int_{-\infty}^{\infty}
\prod_{j \neq i}
d \mu_{j}
{\delta 
(Q\mu_{j} - 
\sum_{l \in \mathbb{N}(i)}}
\Theta (g_{l}-t_{l})
) 
{\rm e}^{-\beta H_{\rm eff}}
}
{
\prod_{i} \sum_{t_{i}=0}^{Q-1} 
\int_{-\infty}^{\infty}
\prod_{i}
d \mu_{i}
{\delta 
(Q\mu_{i} - 
\sum_{l \in \mathbb{N}(i)}}
\Theta (g_{l}-t_{l})
) {\rm e}^{-\beta H_{\rm eff}}
}
\label{eq:localH}
\end{eqnarray}
where $\mathcal{H}_{\rm eff}$ denotes the following 
effective Hamiltonian: 
\begin{eqnarray}
\mathcal{H}_{\rm eff} & = & 
-\sum_{i}
g_{i}\sum_{l \in \mathbb{N}(i)}
\Theta(g_{l}-t_{l})
+
\sum_{i}
\sum_{k,l \in \mathbb{N}(i)}
\Theta(g_{l}-t_{l})\Theta(g_{k}-t_{k})
\end{eqnarray}
In the above expression, 
we omitted the term 
$\sum_{i}g_{i}^{2}$ which is 
independent of the dynamical variable 
$\mbox{\boldmath $t$}$. 
We should keep in mind that 
the normalization 
$\int_{-\infty}^{\infty}
d\mu_{i} P(\mu_{i}: \mbox{\boldmath $g$})=1$ 
is satisfied. 
It also should be noted that  
the pixels are fully connected 
in the window with size $|\mathbb{N}(i)|=Q$. 
Then, the effective Hamiltonian 
is reduced to the decoupled form 
$\mathcal{H}_{\rm eff}=\sum_{i}\mathcal{H}_{\rm eff}^{(i)}$ and we immediately obtain
\begin{eqnarray}
\prod_{j} \sum_{t_{j} \neq t_{i}} 
\int_{-\infty}^{\infty}
\prod_{j \neq i}
d \mu_{j}
\delta 
\left(
Q\mu_{j} - 
\sum_{l \in \mathbb{N}(j)}
\Theta (g_{l}-t_{l})
\right)
\,{\exp}
\left(
-\sum_{j \neq i}\beta H_{\rm eff}^{(j)}
\right) 
& = & {\rm e}^{-(N-1)f} \\
\prod_{i} \sum_{t_{i}}
\int_{-\infty}^{\infty}
\prod_{i}
d \mu_{i}
\delta 
\left(
Q\mu_{i} - 
\sum_{l \in \mathbb{N}(i)}
\Theta (g_{l}-t_{l})
\right) 
\, {\exp}
\left(
-\beta \sum_{i} H_{\rm eff}^{(i)}
\right) & = & {\rm e}^{-Nf}
\end{eqnarray}
by using the saddle point method in 
the limit of $N \to \infty$. 
$f$ is given by $f = -\Phi (\beta,Q: \mbox{\boldmath $g$})$ 
with 
\begin{eqnarray}
\Phi (\beta,Q:\mbox{\boldmath $g$}) =  
\log 
\left\{
\int_{0}^{1}
d\mu 
\int_{-i\infty}^{+i\infty}
\frac{id\hat{\mu}}
{\sqrt{2\pi}}
{\exp}
\left[
-Q\hat{\mu}\mu
-\frac{\beta}{2}
Q^{2}\mu^{2} +
\beta Qg \mu 
+
Q
\log \sum_{t=0}^{Q-1}
{\rm e}^{\hat{\mu} \Theta (g-t)}
\right]
\right\}.
\end{eqnarray}
Then, equation (\ref{eq:localH}) leads to 
\begin{eqnarray}
P_{\beta,Q}(\mu :\mbox{\boldmath $g$}) & = & 
\frac{
{\rm e}^{-\frac{\beta}{2}Q^{2}\mu^{2}+\beta Qg \mu}
\int_{-i\infty}^{+i\infty}
\frac{d\hat{\mu}}
{\sqrt{2\pi}}
{\exp}
\left[
-Q\hat{\mu}\mu
+
Q
\log \sum_{t=0}^{Q-1}
{\rm e}^{\hat{\mu} \Theta (g-t)}
\right]
}
{
\int_{0}^{1}d\mu 
\int_{-i\infty}^{+i\infty}
d \hat{\mu}
\,{\cal B}_{\beta,Q}(\mu,\hat{\mu})
} \\
{\cal B}_{\beta,Q}(\mu,\hat{\mu}) & \equiv & 
\frac{1}{\sqrt{2\pi}}
\exp
\left[
-Q\hat{\mu}\mu 
-\frac{\beta}{2}Q^{2} \mu^{2}
+\beta Q g \mu 
+
Q \log 
\sum_{t=0}^{Q-1}
{\rm e}^{ \hat{\mu} \Theta (g-t)}
\right]. 
\end{eqnarray}
Obviously, the above distribution 
is dependent on 
the original image $\mbox{\boldmath $g$}$ and 
such data-averaged 
distribution is evaluated after slightly 
complicated algebra as follows. 
\begin{eqnarray}
P_{\beta,Q}(\mu) & \equiv & 
\mathbb{E}_{\mbox{\boldmath $g$}} 
[P_{\beta,Q}(\mu :\mbox{\boldmath $g$})]=  
\mathbb{E}_{\mbox{\boldmath $g$}}
\left[
\frac{
\frac{1}{2\pi}\,
{\exp}
\left(-\frac{\beta}{2}
Q^{2} \mu^{2} + \beta Q g \mu 
\right)\, 
\int_{-\infty}^{\infty}
d\hat{\mu}
\cos 
\left[
\mu \hat{\mu}
-
Q \tan^{-1} 
\psi_{Q}(\hat{\mu}) 
\right]
}
{
\int_{-\infty}^{\infty} D\hat{\mu}
\,{\cal Z}_{\beta,Q}(\hat{\mu})
}
\right]
\end{eqnarray}
where 
$\mathbb{E}_{\mbox{\boldmath $g$}}
[\cdots] \equiv 
\int d\mbox{\boldmath $g$}
(\cdots) P(\mbox{\boldmath $g$})$ and 
we defined 
the following functions. 
\begin{eqnarray}
{\cal Z}_{\beta,Q}(\hat{\mu}) & \equiv & 
\xi_{\beta,Q} \cos \Psi_{\beta,Q}(\hat{\mu})
+
\mathcal{F}_{0}(\hat{\mu}) 
\sin \Psi_{\beta,Q}(\hat{\mu}) +
{\rm e}^{-\frac{\beta Q^{2}}{2}}
\left\{
\mathcal{F}_{2}^{(\beta,Q)} (\hat{\mu}) \cos 
\Psi_{\beta,Q}(\hat{\mu})
-
\mathcal{F}_{1}^{(\beta,Q)} (\hat{\mu})
\sin \Psi_{\beta,Q}(\hat{\mu})
\right\} \\
\Psi_{\beta,Q}(\hat{\mu}) & \equiv  & 
Q 
\tan^{-1}
\left\{
\frac{\sum_{t=0}^{Q-1}
{\rm e}^{\beta g \Theta (g-t)} 
\sin \sqrt{\beta} 
\hat{\mu}
\Theta (g-t)}
{\sum_{t=0}^{Q-1}
{\rm e}^{\beta g \Theta (g-t)} 
\cos \sqrt{\beta}
\hat{\mu}
\Theta (g-t)}
\right\} \\
\psi_{Q}(\hat{\mu}) & \equiv & 
\frac{\sum_{t=0}^{Q-1}
\sin (\hat{\mu}/Q)
\Theta (g-t)}
{
\sum_{t=0}^{Q-1}
\cos (\hat{\mu}/Q)
\Theta (g-t)} \\
\xi_{\beta,Q} & \equiv & 
\frac{1}{2} -
H(\sqrt{\beta}Q), \,\,\,\,
H(x) \equiv 
\int_{x}^{\infty}
\frac{dt}{\sqrt{2\pi}}\,
{\rm e}^{-\frac{t^{2}}{2}} \\
\mathcal{F}_{0}(\hat{\mu}) & \equiv & 
\int_{0}^{\hat{\mu}}
\frac{d\lambda}
{\sqrt{2\pi}}\,
{\rm e}^{\frac{\lambda^{2}}{2}},\,\,\,
\mathcal{F}_{1}^{(\beta,Q)} (\hat{\mu}) \equiv  
\int_{0}^{\hat{\mu}}
\frac{d\lambda}
{\sqrt{2\pi}}\,
{\rm e}^{\frac{\lambda^{2}}{2}}
\cos \sqrt{\beta}Q \lambda,\,\,\,
\mathcal{F}_{2}^{(\beta,Q)} (\hat{\mu}) \equiv  
\int_{0}^{\hat{\mu}}
\frac{d\lambda}
{\sqrt{2\pi}}\,
{\rm e}^{\frac{\lambda^{2}}{2}}
\sin \sqrt{\beta}Q \lambda
\end{eqnarray}
As a demonstration, 
we choose uniform grayscale images 
having a single grayscale $g_{0}$, 
namely, 
the distribution of 
$\mbox{\boldmath $G$}$ is given by 
$P(g)=\delta (g-g_{0})$. 
Then, the data-average is easily performed as 
\begin{eqnarray}
P_{\beta,Q}(\mu) & \equiv & 
\mathbb{E}_{\mbox{\scriptsize \boldmath $g$}}[P(\mu :\mbox{\boldmath $g$})]
= P(\mu :g_{0})
\end{eqnarray}
In FIG. \ref{fig:fg_density}, 
we plot the $P_{\beta,Q}(\mu)$ for 
several values of $Q$ and $\beta$ and $g$. 
From this figure, we find that 
as temperature 
$\beta^{-1}$ decreases, the distribution 
changes its shape to 
the delta function in which the peak is 
located at $g_{0}/Q$. 
This means that simulated annealing 
safely finds the ground state of 
the Hamiltonian having the zero energy, namely, 
\begin{eqnarray}
\forall_{i} \,\,\,\,\,\, g_{0}=g_{i} & = & 
\sum_{k \in \mathbb{N}(i)}
\Theta (g_{k}-t_{k}) =s_{i}
\end{eqnarray}
This immediately reads 
$\mu_{i}=g_{0}/Q$ for all index $i$. 
Therefore, for a given image having a single 
grayscale $g_{0}$, 
the distribution 
should converge to 
the delta function 
with the peak at $g_{0}/Q$ if 
the annealing schedule of $\beta$ is appropriate. 
\section{Inverse halftoning as an ill-posed problem}
\label{app:appB}
Here we shall consider the inverse process of halftoning as an ill-posed problem. 
In the inverse halftoning, we should solve 
$\forall_{x,y} \,\,\,
h_{x,y} = \Theta (g_{x,y}-t_{x,y})$ 
with respect to $\mbox{\boldmath $g$}$ for a given 
halftoned image $\mbox{\boldmath $h$}$ and 
the mask $\mbox{\boldmath $t$}$. 
Obviously, 
this problem is ill-posed because 
there are a lot of candidates to satisfy the equations. 
In this Appendix, we analytically evaluate several 
relevant quantities to show the difficulty in finding the solution. 
\subsection{Degree of degeneracy for possible solutions}
\begin{figure}[ht]
\begin{center}
\includegraphics[width=10cm]{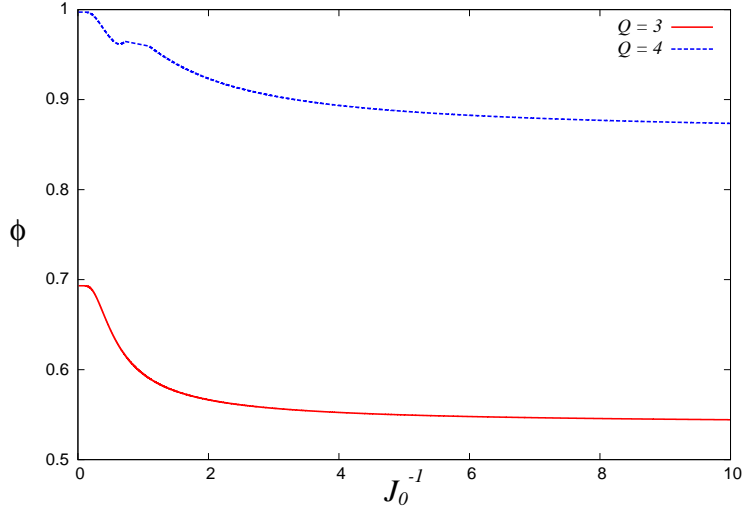}
\end{center}
\caption{\footnotesize 
Logarithm of 
the number of candidates (degeneracy) for 
the solutions $\phi$ as a function 
of $J_{0}^{-1}$ for 
$Q=3$ and $4$.
}
\label{fig:fg_phi}
\end{figure}

We easily assume that the 
number of 
the solutions for a given 
$\mbox{\boldmath $h$}$ and 
$\mbox{\boldmath $t$}$ is 
exponential order, 
however, it is helpful for us 
to evaluate the number more precisely. 
For the purpose, let us use the definition 
introduced in the previous section, namely, 
$i=x+Ly-L, i=1,2,\cdots,L^{2} \equiv N$ 
(conversely, 
$x=(i \mod L), y=[i/L]$). 
Then, the number of 
the candidates for the solution of 
the equations 
$\forall_{i} \,\,\,
h_{i}= \Theta(g_{i}-t_{i})$ 
is evaluated for a 
given a realization of 
the original image and the threshold mask as follows. 
\begin{eqnarray}
\mathcal{N}(\mbox{\boldmath $g$},
\mbox{\boldmath $t$}) & = & 
\prod_{i=1}^{N}
\left\{
(Q-1-t_{i})
\Theta (g_{i}-t_{i})
+t_{i} 
\Theta (t_{i}-g_{i}) 
\right\}
\end{eqnarray}
As the number $\mathcal{N}$ seems to be an exponential 
order object, we might rewrite the data average of 
the logarithm of the number as a self-averaging quantity  
\begin{eqnarray}
\mathbb{E}_{\mbox{\scriptsize \boldmath $g$},\mbox{\scriptsize \boldmath $t$}}
[\log \mathcal{N}(\mbox{\boldmath $g$},
\mbox{\boldmath $t$})]
& = &  
\mathbb{E}_{\mbox{\scriptsize \boldmath $g$},\mbox{\scriptsize \boldmath $t$}}
\left[
\log 
\prod_{i=1}^{N}
\left\{
(Q-1-t_{i})
\Theta (g_{i}-t_{i})
+t_{i}
\Theta (t_{i}-g_{i}) 
\right\}
\right]
\end{eqnarray}
where we defined the expectation $\mathbb{E}_{\mbox{\scriptsize \boldmath $g$},
\mbox{\scriptsize \boldmath $t$}}[\cdots]$ by 
\begin{eqnarray}
\mathbb{E}_{\mbox{\scriptsize \boldmath $g$},
\mbox{\scriptsize \boldmath $t$}}[\cdots] & \equiv & 
\int \int P_{J_{0}}(\mbox{\boldmath $g$})
P(\mbox{\boldmath $t$})
(\cdots)d 
\mbox{\boldmath $g$}\,
d\mbox{\boldmath $t$},\,\,\,
d\mbox{\boldmath $g$} \equiv 
\prod_{i=1}^{N} 
dg_{i},\,
d\mbox{\boldmath $t$} \equiv 
\prod_{i=1}^{N} 
dt_{i}.
\end{eqnarray}
Therefore, the above averages could 
be carried out for a specific choice of 
the $P(\mbox{\boldmath $g$})$ and 
$P(\mbox{\boldmath $t$})$.
For distributions of the threshold mask, 
we suppose that 
in each mask with size 
$Q \times Q$ ($Q \ll N$), 
each component $t_{i}$
takes a value among $0,\cdots,Q-1$ grayscales 
with equal probability 
$1/Q$. 
Then, the $P(\mbox{\boldmath $t$})$ is 
reduced to the product of the effective 
single site distribution as 
\begin{eqnarray}
P(\mbox{\boldmath $t$})= \prod_{i=1}^{N} P(t_{i}) & = & 
\frac{1}{Q} 
\prod_{i=1}^{N} \sum_{l_{i}=0}^{Q-1}
\delta_{l_{i},t_{i}}.
\end{eqnarray}
On the other hand, 
as a distribution of 
the original image, we consider 
snapshots from the infinite-range ferromagnetic Q-Ising model, 
that is, 
\begin{eqnarray}
P_{J_{0}}(\mbox{\boldmath $g$}) & = & 
\frac{\exp (-\mathcal{H})}
{Z_{0}},\,\,\,\,
\mathcal{H} = 
\frac{J_{0}}{2N} 
\sum_{ij}
(g_{i}-g_{j})^{2} 
\label{eq:original_ham}
\end{eqnarray}
where $Z_{0}$ is a normalization constant 
for the probability $P_{J_{0}}(\mbox{\boldmath $g$})$. After simple algebra, we find 
that the $P_{J_{0}}(\mbox{\boldmath $g$})$ is 
rewritten such as 
$P_{J_{0}}(\mbox{\boldmath $g$}) = 
\prod_{i=1}^{N} 
P_{J_{0}}(g_{i})$ with 
the following effective 
single site 
distribution: 
\begin{eqnarray}
P_{J_{0}}(g_{i}) & = & 
\frac{\exp (-J_{0}g_{i}^{2} + 
2J_{0} m_{0} g_{i})}
{\sum_{g_{i}=0}^{Q-1}
\exp (-J_{0}g_{i}^{2} + 
2J_{0} m_{0} g_{i})
}, \,\,\,\,
m_{0} = 
\frac{\sum_{g=0}^{Q-1} g\, \exp (-J_{0}g^{2} + 
2J_{0} m_{0} g)}
{\sum_{g=0}^{Q-1}
\exp (-J_{0}g^{2} + 
2J_{0} m_{0} g)
}
\label{eq:original_mag}
\end{eqnarray}
where $m_{0}=(1/N)\sum_{i=1}^{N}g_{i}$ denotes 
the magnetization 
for the system of 
original grayscale images 
described by the Hamiltonian (\ref{eq:original_ham}). 
For these probability distributions, 
we have the function $\phi (J_{0},Q) \equiv 
N^{-1}
\mathbb{E}_{
\mbox{\scriptsize \boldmath $g$},
\mbox{\scriptsize \boldmath $t$}}
[\log \mathcal{N}(\mbox{\boldmath $g$},
\mbox{\boldmath $t$})]$ explicitly as 
\begin{eqnarray}
\phi(J_{0},Q) & = & 
\sum_{g,t=0}^{Q-1}  
P_{J_{0}}(g)P(t) 
\log 
\left\{
(Q-1-t)\Theta (g-t) + 
t\Theta (t-g)
\right\} \nonumber \\
\mbox{} & = &  
\frac{
\sum_{t=0}^{Q-1}
\sum_{l=0}^{Q-1} 
\delta_{l,t}
\sum_{g=0}^{Q-1}
{\rm e}^{-J_{0}g^{2} + 
2J_{0}
m_{0}g}
\log 
\left\{
(Q-1-t)\Theta (g-t) + 
t\Theta (t-g)
\right\}
}
{
Q
\sum_{g=0}^{Q-1}
{\rm e}^{-J_{0}g^{2} + 
2J_{0}
m_{0}g}}. 
\end{eqnarray}
Therefore, 
the number of the solutions for 
the equations $\forall_{i}\,\,\,
h_{i} =\Theta (g_{i} -t_{i})$ is 
estimated as $\sim \exp(N\phi (J_{0},Q))$. 
In FIG. \ref{fig:fg_phi}, 
we plot the function $\phi(J_{0},3)$ and 
$\phi(J_{0},4)$. 
From this figure, 
we find that 
the $\phi$ is a monotonically decreasing 
function as $J_{0}^{-1}$ decreases, 
however, the $\phi$ remains a finite value even 
if $J_{0}^{-1} \to \infty$. 
This means that the number of 
the candidates for the solutions 
of the inverse halftoning is always exponential order. 
Therefore, 
we need some systematic approach to 
solve this type of the ill-posed problems.  
Namely, we need to 
introduce the ferromagnetic prior $P_{J}(\mbox{\boldmath $\sigma$})$ to 
compensate the lack of information as we discussed in Sec. \ref{sec:INV}. 
\subsection{Mutual information}
We next consider the difficulties of 
retrieving original grayscale images 
from a slightly different point of view. 
Here we calculate the mutual information 
between the original image $\mbox{\boldmath $g$}$ and 
the halftone image $\mbox{\boldmath $h$}$. 
From the definition of 
the mutual information, we should evaluate 
$I(\mbox{\boldmath $g$}: 
\mbox{\boldmath $h$})= 
H(\mbox{\boldmath $h$})- 
H(\mbox{\boldmath $h$}|\mbox{\boldmath $g$})$, 
where we defined the entropy $H(\mbox{\boldmath $h$})$ 
and the conditional entropy $H(\mbox{\boldmath $h$}|\mbox{\boldmath $g$})$ 
as $H(\mbox{\boldmath $h$})=
-\sum_{\mbox{\scriptsize \boldmath $h$}}
P(\mbox{\boldmath $h$})\log P(\mbox{\boldmath $h$}), 
H(\mbox{\boldmath $h$}|\mbox{\boldmath $g$})=
-\sum_{\mbox{\scriptsize \boldmath $g$}}P_{J_{0}}(\mbox{\boldmath $g$})
\sum_{\mbox{\scriptsize \boldmath $h$}}
P_{t}(\mbox{\boldmath $h$}|\mbox{\boldmath $g$})\log 
P_{t}(\mbox{\boldmath $h$}|\mbox{\boldmath $g$})$, 
respectively. 
It should be noted that we used 
\begin{eqnarray}
P_{t}(\mbox{\boldmath $h$}|
\mbox{\boldmath $g$}) & = & 
\prod_{i}\delta (h_{i},\Theta (g_{i}-t_{i}))
\end{eqnarray}
and $P(\mbox{\boldmath $h$})=
\sum_{\mbox{\scriptsize \boldmath $g$}}P_{t}(\mbox{\boldmath $h$}|
\mbox{\boldmath $g$})P_{J_{0}}(\mbox{\boldmath $g$})$. 

\begin{figure}[ht]
\begin{center}
\includegraphics[width=10cm]{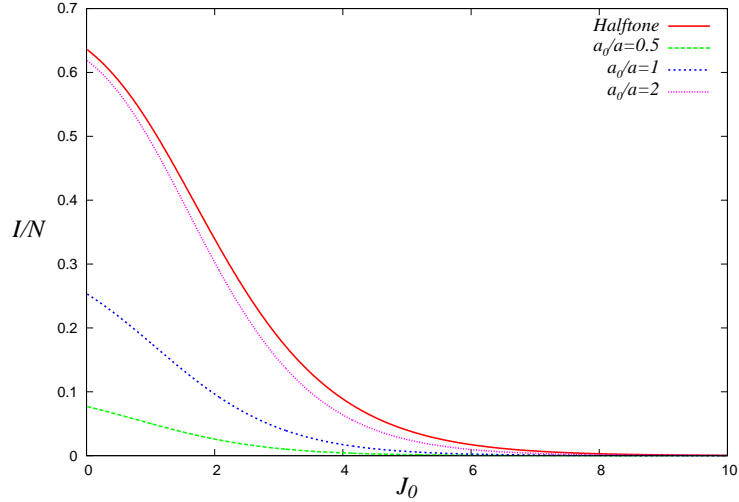}
\end{center}
\caption{\footnotesize 
The mutual information per pixel as a function of $J_{0}$. 
For the halftoning process, we choose 
$Q=3$ and $t_{0}=1$. 
The mutual information 
for image restoration  under Gaussian noise is also 
plotted for the signal to noise ratio $a_{0}/a=0.5,1$ and $2$. 
}
\label{fig:fg_mutinfo}
\end{figure}

For the infinite-range model and 
constant mask $t_{i}=t_{0}\,\,\,
\forall_{i}$, these 
entropies per are calculated analytically as 
\begin{eqnarray}
\frac{H(\mbox{\boldmath $h$})}{N} & = & 
-
\sum_{h=0,1}
\sum_{g=0}^{Q-1}
\left\{
\frac{{\rm e}^{-J_{0}g^{2}+2J_{0}m_{0}g}
\delta (h,\Theta(g-t_{0}))}
{Z_{J_{0}}}
\right\}
\log 
\left\{
\frac{\sum_{g=0}^{Q-1}
{\rm e}^{-J_{0}g^{2}+2J_{0}m_{0}g}
\delta (h,\Theta(g-t_{0}))}
{Z_{J_{0}}}
\right\} \\
\frac{H(\mbox{\boldmath $h$}|\mbox{\boldmath $g$})}{N} & = & 
-\sum_{g=0}^{Q-1}
\left\{
\frac{
{\rm e}^{-J_{0}g^{2}+2J_{0}m_{0}g}}
{Z_{J_{0}}}
\right\}
\sum_{h=0,1}
\delta (h,\Theta (g-t_{0})) 
\log 
\delta (h,\Theta (g-t_{0})) =0
\end{eqnarray}
where we defined 
$Z_{J_{0}} \equiv 
\sum_{g=0}^{Q-1}
{\rm e}^{-J_{0}g^{2}+2J_{0}m_{0}g}$ 
and $m_{0}$ is a solution 
of the equation (\ref{eq:original_mag}). 
We should keep in mind that the `channel' of the halftoning process is 
completely deterministic. As the result, the conditional entropy 
$H(\mbox{\boldmath $h$}|\mbox{\boldmath $g$})$ is identically zero. 
In FIG. \ref{fig:fg_mutinfo}, 
we plot the mutual information per pixel 
$I(\mbox{\boldmath $g$}: \mbox{\boldmath $h$})/N$ 
as a function of $J_{0}$. 
To compare 
halftoning process by the dither method 
with the degraded process by a Gaussian noise 
with mean $a_{0}g$ and the variance $a^{2}$, 
we also calculate the mutual information 
per pixel for the Gaussian noise. 
We choose the same distribution 
$P_{J_{0}}(\mbox{\boldmath $g$})$ as the halftone case. 
We immediately have 
\begin{eqnarray}
\frac{H(\mbox{\boldmath $h$})}{N} & = & 
-
\frac{1}{\sqrt{2\pi} a Z_{J_{0}}}
\int_{-\infty}^{\infty}
dh 
\sum_{g=0}^{Q-1}
{\rm e}^{-J_{0}g^{2}+2J_{0}m_{0}g- 
\frac{(h-a_{0}g)^{2}}{2a^{2}}}
\log 
\left\{
\frac{
\sum_{g=0}^{Q-1}
{\rm e}^{-J_{0}g^{2}+2J_{0}m_{0}g- 
\frac{(h-a_{0}g)^{2}}{2a^{2}}}
}
{\sqrt{2\pi} a Z_{J_{0}}}
\right\} \\
\frac{H(\mbox{\boldmath $h$}|\mbox{\boldmath $g$})}{N} & = & 
\frac{1}{2} + \log (\sqrt{2\pi} a).
\end{eqnarray}
In FIG. \ref{fig:fg_mutinfo}, 
we plot the mutual information per pixel for 
the Gaussian noise as a function of 
$J_{0}$ for various cases of 
the signal-noise ratio, $a_{0}/a$. 

\end{document}